\documentclass[acmlarge]{acmart}
\AtBeginDocument{%
  }

\setcopyright{acmlicensed}
\copyrightyear{2018}
\acmYear{2018}
\acmDOI{XXXXXXX.XXXXXXX}
\acmConference[Conference acronym 'XX]{Make sure to enter the correct
  conference title from your rights confirmation email}{June 03--05,
  2018}{Woodstock, NY}
\acmISBN{978-1-4503-XXXX-X/2018/06}

\usepackage{pifont}

\newcommand{\cmark}{\ding{51}}

\usepackage{multirow}
\usepackage{xcolor}
\usepackage{tabularx}
\usepackage{booktabs}
\usepackage{enumitem}
\usepackage{url}
\usepackage{longtable}
\usepackage{pdflscape}
\usepackage{graphicx}
\usepackage{subcaption}
\usepackage{tablefootnote}

\usepackage{array}
\usepackage{ragged2e}
\usepackage{colortbl}

\newcolumntype{Z}[1]{>{\RaggedRight\arraybackslash\hspace{0pt}}p{#1}}

\newcommand{\layerdesc}[1]{\textcolor{gray}{\footnotesize #1}}

\begin{document}

\title[LaughAnchor]{Reconstruction and Reflection of Positive Experiences through Resurfacing Laughter-indexed Everyday Moments}


\author{Jun Fang}
\authornote{Both authors contributed equally to this research.}
\affiliation{%
  \institution{Tsinghua University}
  \city{Beijing}
  \country{China}
}
\email{fangy23@mails.tsinghua.edu.cn}
\orcid{0009-0001-2614-8674}
\author{Jiajin Li}
\authornotemark[1]
\authornote{Intern at the time of this work.}
\affiliation{%
  \institution{Tsinghua University}
  \city{Beijing}
  \country{China}
}
\email{lijiajin0516@gmail.com}
\orcid{0009-0001-5723-5383}

\author{Yuntao Wang}
\authornote{Corresponding author.}
\orcid{0000-0002-4249-8893}
\affiliation{%
  \institution{Tsinghua University}
  \city{Beijing}
  \country{China}
}
\email{yuntaowang@tsinghua.edu.cn}

% intern
\author{Kexin Miao}
\authornotemark[2]
\affiliation{%
  \institution{Tsinghua University}
  \city{Beijing}
  \country{China}
}
\email{kexinmiao22@outlook.com}
\orcid{0009-0003-5844-0033}

\author{Susu Wang}
\affiliation{
  \institution{Beijing University of Chemical Technology}
  \city{Beijing}
  \country{China}
}
\email{19156131236@163.com}
\orcid{0009-0004-6002-7085}

\author{Xiaoyu Xie}
\affiliation{
  \institution{Hunan University}
  \city{Changsha}
  \country{China}
}
\email{2864209248@qq.com}
\orcid{0009-0003-5562-8137}

\author{Kaixin Ji}
\orcid{0000-0002-4679-4526}
\affiliation{%
  \institution{Tsinghua University}
  \city{Beijing}
  \country{China}
}
\email{ji-kx@mail.tsinghua.edu.cn}

\author{Yuanchun Shi}
\orcid{0000-0003-2273-6927}
\affiliation{%
  \institution{Tsinghua University}
  \city{Beijing}
  \country{China}
}
\email{shiyc@tsinghua.edu.cn}

\renewcommand{\shortauthors}{Fang et al.}

\begin{abstract}

\begin{figure}[h]
  \centering
  \includegraphics[width=\textwidth]{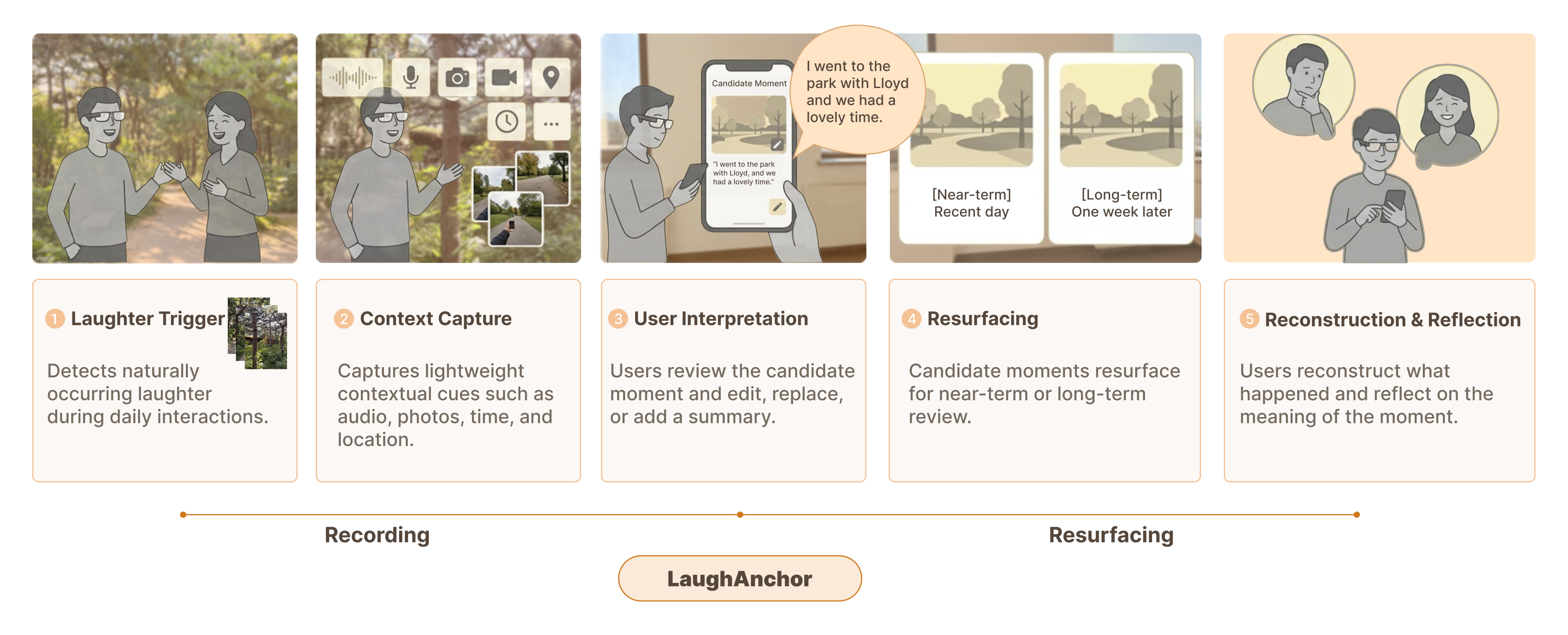}
  \caption{Overview of how \textit{LaughAnchor} uses naturally occurring laughter to construct contextualized personal records that users can interpret and curate. During participant-initiated recording, laughter serves as a sparse affective index for assembling aligned contextual cues into candidate Moment Cards. The system later resurfaces retained cards at near-term and long-term intervals to support reconstruction and reflection.}
  \label{fig:teaser}
\end{figure}

Positive everyday moments often escape deliberate recording, while continuous self-tracking can generate extensive records that are difficult to revisit. We explore laughter as a naturally occurring, sparse index for constructing contextualized personal records to support later reconstruction and reflection. A formative study with 12 participants characterized laughter as an affective but semantically incomplete index and informed \textit{LaughAnchor}, a mobile and wearable self-tracking system. During participant-initiated recording, the system assembles detected laughter and aligned context into candidate moments for later reconstruction and reflection, with layered context disclosure, user-controlled curation, and near-term and long-term resurfacing. In a three-week field deployment with 12 participants, passive indexing preserved moments they considered unlikely to record deliberately but valued retrospectively. During resurfacing, participants attributed affective re-experiencing to laughter and used additional context both to reconstruct episodes and to explore already-recalled experiences. Across moments and reviews, resurfacing supported rediscovery and broader awareness of relationships, routines, and emotional states. These findings inform self-tracking designs that use sparse affective indices to organize contextual records for reconstruction and reflection, while keeping interpretation and retention under user control.

\end{abstract}




%   \centering
%   \includegraphics[width=\textwidth]{figure/Teaser.png}
%   \caption{Overview of how \textit{LaughAnchor} uses naturally occurring laughter to construct contextualized personal records that users can interpret and curate. During participant-initiated recording, laughter serves as a sparse affective index for assembling aligned contextual cues into candidate Moment Cards. The system later resurfaces retained cards at near-term and long-term intervals to support reconstruction and reflection.}
%   \Description{A five-stage overview of the LaughAnchor workflow, including laughter trigger, context capture, user interpretation, resurfacing, and reconstruction and reflection.}

\received{20 February 2007}
\received[revised]{12 March 2009}
\received[accepted]{5 June 2009}

\maketitle

% 正文
\section{Introduction}
\label{intro}
% 一页

% LaughAnchor 研究的不是如何换一种形式呈现笑声，而是如何以笑声为稀疏情感索引，构造可被语境化、由用户解释和策展、并在日常再次使用的个人记录。

% part1 -- background and significance [为什么重要] -- 为什么 daily positive moments 值得系统支持，但现有记录方式不好处理
% 1. Positive moments are valuable to revisit, but are fleeting and rarely worth deliberate capture:
% self-track systems 帮助用户记录并反思生活经历。回看 positive moments 可以支持情绪调节、自我理解和 wellbeing (cite Isaacs et al., 2013, Echoes from the Past; Avrahami et al., 2020, Celebrating Everyday Success)。
% 2. daily positive moments 往往短暂、无计划、过于普通，用户不会每次positive moments都主动及时记录。
% 滞后会导致信息遗漏或遗忘记录，回忆中很难重温情绪，很难获得心理需求的感知与共鸣 \cite{hollis2017does}。然而，被动式连续self-track systems continuously record user’s signals\cite{hollis2017does, mann1996wearable, jiang2019memento}，continuous lifelogs overwhelm review (cite{Pixel Memorieschi2025})
% 传统方式下positive moments的resurface，形成了记录选择性问题 -- 主动记录具有较强的个人相关性，但容易遗漏普通而短暂的经历；持续记录覆盖更广，却产生大量难以有效回看的材料，how to preserve lightweight traces of potentially valuable positive moments that remain manageable and meaningful to revisit.

Everyday life is punctuated by brief positive moments, such as an unexpected joke in conversation, a playful exchange with a friend, or a moment of warmth. 
Although often ordinary and easily overlooked, these moments can later reveal sources of enjoyment, connection, and meaning within everyday routines.
Resurfacing such experiences can support positive affect~\cite{fredrickson2013love}, autobiographical reflection~\cite{isaacs2013echoes}, and a more nuanced understanding of routine life~\cite{avrahami2020celebrating}. 

Self-tracking systems increasingly help people preserve and reflect on lived experiences beyond numerical records~\cite{kato2024tippy,jang2025journey,mcduff2012affectaura,konrad2016technology}. 
Yet their ability to support later reflection depends on what enters the record in the first place.
Existing approaches face a fundamental capture trade-off: deliberate recording requires people to recognize an experience as worth preserving while it is happening, and may therefore miss brief or socially absorbing moments, as well as those whose value becomes apparent only later~\cite{mols2014making}. Continuous capture postpones this judgment, but produces large streams of contextual material that users must later segment and review~\cite{hollis2017does,jiang2019memento,elagroudy2025pixel}.
The challenge is therefore to preserve sparse traces of potentially valuable moments without requiring their value to be known in advance, while keeping these traces manageable for later reconstruction and reflection.
Laughter may serve as a promising cue for addressing this trade-off.
It occurs naturally as a temporally localized vocal event in everyday interaction~\cite{vettin2004laughter,ryokai2018capturing}, providing sparse anchors within longer periods of activity.
Its acoustic form and surrounding audio can carry cues about the emotional tone, atmosphere, and social interaction of an episode~\cite{sauter2010perceptual,scott2014social,bryant2016detecting}.
Although not every positive moment involves laughter, audible laughter commonly communicates amusement or positive affect across cultural contexts~\cite{sauter2010cross,gendron2014cultural,bryant2022laughter}.
Yet laughter is affectively informative but semantically incomplete: it may indicate that an affectively salient episode occurred without specifying what happened, why it mattered, or whether it is worth preserving.
We therefore treat laughter not as a label for positive experience, but as a sparse affective index into the surrounding episode. We define a \textit{laughter-indexed moment} as a candidate moment anchored by one or more detected laughter bouts and scaffolded with bounded contextual cues for later reconstruction and reflection.

% We take the surrounding episode as the unit of design, examining how laughter-indexed candidates can be reconstructed, interpreted, and revisited as personally meaningful records.

Prior work has explored how captured laughter can be preserved and revisited through tangible representations~\cite{ryokai2018capturing} and visualizations of its temporal, spatial, and social aspects~\cite{yang2022exploring}. 
These studies established the evocative value of recorded laughter. Laughter has also been used to index video clips of surrounding activity~\cite{fukumoto2013smile}. 
A laughter trace may convey the emotional tone of an experience without fully specifying what happened. We therefore take the surrounding episode as the unit of design, using laughter as an affective index that can be progressively grounded through contextual cues and user interpretation. Layered context disclosure, revisable user curation, and repeated resurfacing implement this approach.
Motivated by this, we propose two research questions:

\begin{itemize}
    \item[\textbf{RQ1:}] \textit{How can passively indexed laughter be transformed into reconstructable candidate moments of potentially positive everyday experience?}
    \item[\textbf{RQ2:}] \textit{How do people engage with laughter-indexed moments when they are resurfaced in everyday life?}
\end{itemize}

To address these RQs, we conducted a formative study with 12 participants, combining interviews, natural recording, and think-aloud reviews of their own laughter-indexed moments. The study characterized laughter as affectively expressive but semantically incomplete, with contextual cues supporting episode reconstruction and users determining personal significance. These findings informed \textit{LaughAnchor}, a mobile and wearable system that assembles detected laughter and aligned context into candidate moments during participant-initiated recording. It supports layered context disclosure, user curation, and near-term and long-term resurfacing. We then conducted a three-week field deployment with 12 participants to examine how these records complemented existing practices and supported reconstruction and reflection.
The deployment showed that laughter could provide an affective entry point even without complete episodic recall, while contextual cues supported both reconstruction and further exploration after an episode had already been recalled.
This work makes three contributions:

\begin{itemize}
    \item We contribute a design approach, grounded in a formative study, for using laughter as a sparse affective index to construct contextualized personal records. It identifies how aligned contextual cues support episode reconstruction and how users assess the reflective value of the resulting moments.

    \item We present \textit{LaughAnchor}, a mobile and wearable implementation that connects laughter indexing with contextual moment cards, layered disclosure, revisable user curation, and near-term and long-term resurfacing.

    \item We provide an empirical account from a three-week field deployment ($N=12$), showing how \textit{LaughAnchor} complements existing recording practices and supports affective engagement without complete episodic recall, continued contextual exploration after and beyond recall, and reflection across accumulated moments.
\end{itemize}

\section{Background and Related Work}
% 一页半

\subsection{Self-Tracking for Positive Everyday Moments}

% | Paragraph 1 | 定义范围并建立分类 | self-tracking 如何记录 positive moments？记录选择由谁、在何时完成？ |
% self-tracking 不只是记录数字，也可以帮助人回看日常经验；positive moments 的回看可能支持积极感受、自我理解与日常生活评价。
% 现有工作 Deliberate capture；Continuous or passive capture。围绕三个gap，从capture到reflection，一类工作通过丰富 context、timeline、calendar、summary 支持 retrieval 和 reminiscence；一类工作通过 prompting、narrative、annotation 或 user interpretation 支持 meaning-making。

Self-tracking systems increasingly capture lived experiences, including emotions, activities, and personally meaningful events, alongside numerical states and behaviors~\cite{staahl2009experiencing,mcduff2012affectaura,konrad2016technology}. Positive everyday moments merit attention because they are often brief, ordinary, and easily overlooked, while negative information attracts more attention and remains cognitively accessible~\cite{mols2014making,baumeister2001bad,ledgerwood2014sticky}. Capturing and resurfacing positive experiences can support affective re-experiencing, highlight everyday sources of enjoyment and warmth, and provide material for reflection, helping people appreciate experiences that have already occurred~\cite{kato2024tippy,isaacs2013echoes,avrahami2020celebrating}.

% | Paragraph 2 | Deliberate capture | 用户在记录前或记录时完成选择，因此 personal relevance 较高，但可能遗漏当下未被识别为值得记录的 moment | continuous capture | 系统先扩大覆盖，用户在之后回看时完成选择，因此降低即时操作，但将负担转移到 review、filtering 和 privacy management |
Existing approaches distribute effort and judgment differently. Deliberate approaches require users to select and externalize experiences, while passive and continuous approaches shift capture toward sensing and computational organization.
Prior systems of deliberate self-tracking have supported photos of small positive moments~\cite{kato2024tippy}, textual records of positive experiences~\cite{isaacs2013echoes}, records of positive workplace events~\cite{avrahami2020celebrating}, and prompts about past proud moments~\cite{jang2025journey}. 
However, they require users either to recognize an experience as record-worthy while it unfolds, or to retrieve and articulate it later.
Brief, ordinary, or socially absorbing moments may remain unrecorded, especially when their value emerges only in retrospect~\cite{mols2014making}. 
Passive and sensor-augmented approaches redistribute effort by collecting contextual or affective traces with less in-the-moment input. \citet{staahl2009experiencing} and \citet{mcduff2012affectaura} combine sensed information with personal media and temporal records to support later interpretation of emotional experiences, while continuous self-tracking broadens coverage by retaining larger streams of everyday context~\cite{hollis2017does,jiang2019memento}. 
These systems can preserve experiences that were not consciously selected as they occurred. However, active recording can itself provide an opportunity for reflection~\cite{rivera2017introducing}. Reducing capture effort does not eliminate selection or interpretation. Instead, effort shifts to later review. Users must identify relevant episodes, distinguish meaningful moments from low-value material, and manage extensive or privacy-sensitive data. Event segmentation, timelines, summaries, and generated memory prompts can make these archives more manageable, but they still require users or systems to define event boundaries~\cite{elagroudy2025pixel}.
Moreover, large amounts of passively captured, fine-grained context do not automatically yield personally meaningful records because the value of positive moments remains subjective~\cite{lee2025designing}.

% | Paragraph 3 | 综合 trade-off 并定位本文 | 是否存在一种在 capture 时就用稀疏信号划定 candidate、但不自动决定意义的中间路径？ 
% 在“费力的主动记录”与“信息过载的连续记录”之间缺少一种低负担的方式。
% 最后说，Existing practices miss or overwhelm everyday moments，所以本文...
% These approaches expose a trade-off among recording effort, information volume, and event boundaries. We explore an intermediate approach using a sparse, naturally occurring signal to index manageable candidate moments without assigning meaning or value in advance. We investigate laughter as such an index to reduce in-the-moment recording effort while keeping reconstruction and reflection user-led. Laughter-indexed moments can later be resurfaced with layered contextual cues, while users decide whether to retain them and allow future resurfacing.

These approaches expose a trade-off among recording effort, information volume, and event boundaries. \textit{LaughAnchor} explores an intermediate approach, using naturally occurring laughter as a sparse index to reduce in-the-moment recording effort and organize contextualized candidate moments for later reconstruction and user interpretation.

\subsection{Laughter as an Affective yet Ambiguous Index}

% 1. 传统 emotion tracking 多依赖文字、照片，hci 探索手势、面部运动等nonverbal traces，和emotional moments紧密相连，增强情绪重温，同时从ubiquitous computing角度易于capture。这些方式说明情绪经验并不只能通过语言文字图片记录，可以通过副语言信息等trace。

Affective experiences leave traces beyond deliberately authored text and photos. HCI research has examined physiological signals~\cite{niforatos2015pulsecam} and paralinguistic information, such as gestures~\cite{luo2024emotion}, facial movements~\cite{yan2022emoglass} and acoustic features~\cite{mcduff2012affectaura}. These modalities can reduce reliance on explicit self-description and preserve transient qualities that static records may omit. However, such traces rarely identify the underlying experience or its personal significance on their own. Their interpretation usually depends on the surrounding activity, people, location, and other contextual information.

% 2. laughter 是一类特别的 trace，它具有特别的属性：社会学、心理学；

Within this broader design space, laughter provides a distinctive paralinguistic index. It occurs naturally in discrete bouts within ongoing everyday interaction~\cite{vettin2004laughter}. The acoustic characteristics of laughter can convey emotional tone, while its timing and form are closely shaped by social interaction~\cite{sauter2010perceptual,scott2014social}. Even brief instances of shared laughter can convey affiliation between speakers~\cite{bryant2016detecting}. Laughter commonly communicates amusement and positive affect. Its broad affective valence may be more consistent across cultures than specific emotional interpretations~\cite{gendron2014cultural,bryant2022laughter}. These properties make laughter a plausible index of potentially positive and socially meaningful experiences. Nevertheless, laughter alone does not establish positivity, personal importance, or reflective value.
We therefore treat it as a candidate anchor for potentially positive moments.

% 3. 现有 laughter HCI work 证明 potential，但没有完成 signal-to-moment transformation

% p.s. （两三句说明技术可行。上次wyt说，appendix要放我们跑模型的results）audio-based laughter detection 支持真实场景 prototype；wearable sensing 已被用于捕捉表情、声音和情绪相关信号；本文将笑声检测视作一项支撑性模块，而非本研究的核心技术贡献。

Advances in audio-based methods have enabled laughter detection in noisy recordings and wearable sensing settings, although performance remains sensitive to environmental noise and differences between training and everyday conditions~\cite{gillick2021robust,hagerer2018robust}. HCI research has also established the potential of personal laughter as material for preservation and reflection. \citet{ryokai2018capturing} captured naturally occurring laughter and explored tangible representations that transformed laughter sounds into concrete reminders for preservation, interaction, and reflection. \citet{yang2022exploring} subsequently examined visualizations of the temporal, spatial, and social dimensions of personal laughter, exploring how participants engaged with these representations and related them to personal memories, emotions, and relationships. Laughter Map~\cite{shigi2023laughter, shigi2024does} further combined recordings of laughter and surrounding conversation with geographic visualization to support the recall of pleasant experiences. 

% These studies establish the evocative value of personal laughter and the importance of personal and interactional context. \citet{ryokai2018capturing} also identified a tension between attending to laughter itself and preserving context for later recall. Building on these insights, we examine how passively captured laughter can be organized into bounded candidate moments, what contextual cues support reconstruction, and how users assess their reflective value and decide whether they should be retained or resurfaced. \textit{LaughAnchor} combines layered context disclosure and user-controlled curation to investigate reconstruction and reflection during near-term and long-term resurfacing.

~\citet{ryokai2018capturing} also identified a tension between attending to laughter itself and preserving context, proposing background context archiving as a future design direction. 
Our work therefore investigates how laughter-indexed candidates become contextualized personal records that users reconstruct, interpret, and later resurface. 
We examine how people draw on contextual materials before and after recalling an episode, and how accumulated moments and reviews inform their understanding of everyday experiences. \textit{LaughAnchor} implements this approach through layered context disclosure, user-controlled curation, and near-term and long-term resurfacing.

\subsection{Resurfacing for Reconstruction and Reflection}
\label{chap:re_three}

%Remembering is ……？memory retrival is ?
Autobiographical remembering reconstructs past experience. It does not simply replay a stored record. People combine cues, general knowledge, and event-specific details to rebuild an experienced scene~\cite{rubin2015event,van2008informing}. Memory systems can support this process by moving users from familiarity or coarse recognition toward richer reconstruction~\cite{sellen2007life}. Contextual cues can ground otherwise fragmented traces, but excessive information may increase cognitive burden, expose sensitive content, or constrain interpretation~\cite{staahl2009experiencing,bellotti1993design}. Memory systems should scaffold user-led reconstruction without taking over the interpretive process.

Reflection is distinct from reconstruction. Identifying what happened does not necessarily make a moment meaningful. Reflection involves intellectual and affective engagement through which people develop new understanding or appreciation of an experience~\cite{fleck2010reflecting}. A recognizable moment may remain ordinary or unwanted, whereas an incomplete memory may still prompt insight into a relationship, a period of life, or a current emotional state. Systems should therefore preserve user authority over meaning and retention.

We use \textit{resurfacing} to describe the system-mediated re-presentation of a previously captured moment at a later time or in a later context. Research on proactive, location-based resurfacing suggests that it can bring forward everyday memories that users would otherwise be unlikely to recall~\cite{mcgookin2019reveal}. However, resurfacing creates an opportunity for reconstruction and reflection rather than guaranteeing either process. Whether these processes occur depends on timing, context, and user control.
A recent systematic review proposes a cue-centered framework covering cue generation, augmentation, interaction, and sharing in reminiscence technologies~\cite{zhang2026memory}. Within this design space, \textit{LaughAnchor} supports engagement with laughter-indexed moments through layered context disclosure and user-controlled resurfacing. We examine how episode reconstruction relates to affective engagement, continued exploration, and reflection on the personal meaning of these experiences.
% Prior work has treated cue-supported memory reconstruction~\cite{sellen2007life,van2008informing}, reflection on experience~\cite{staahl2009experiencing}, and proactive memory resurfacing~\cite{mcgookin2019reveal} as distinct concerns. 
% \textit{LaughAnchor} brings these concerns together through context-scaffolded resurfacing that supports user-led reconstruction and reflection, with users deciding what each moment means and whether it should be retained or resurfaced again.

% 1. remembering is reconstructive, from know to remember. cue 可以帮助恢复details，但是本身停留在 knowing / familiarity。

% 2. Reconstruction 中，context cues 应支持用户主导的重构，而不是限定或者替代用户的解读，even when a moment is recognizable, its value and meaning remain user-authored.

% 最后说，Existing work has not integrated reconstruction, reflection into an end-to-end process to support meaning attribution, curation, and resurfacing laughter-indexed moments，所以本文 xxx

\section{Apparatus and Laughter-Indexing Pipeline}
\label{sec:capture-pipeline}
% 半页

% 本章
% sensor stream → laughter detection → laughter bout → candidate laughter episode

% Formative Study
% candidate laughter episode → context selection → context layers
% → reconstruction / reflection probes

% System Design
% design considerations → final presentation, curation, and resurfacing mechanisms

% aiget 在Apparatus部分并没有写具体每个环节的implementation & api，在formative Apparatus部分也没写api。
% 功耗前后可能不同，要不类似aiget，直接都不报道，要不对应part提一句。
% Laughter Capture Workflow不涉及context info获取方式，因为还没讲layer。layer选取逻辑和获取方式通过简短1/4页的表达 + 表，不必赘述，不用写example。因为system outline大概率会标/有图。

The formative study and field deployment used the same core sensing architecture and laughter-indexing logic, with parameter settings adapted to each study. This section summarizes the shared apparatus and processing workflow.

\begin{figure}[h]
    \centering
    \includegraphics[width=.4\linewidth]{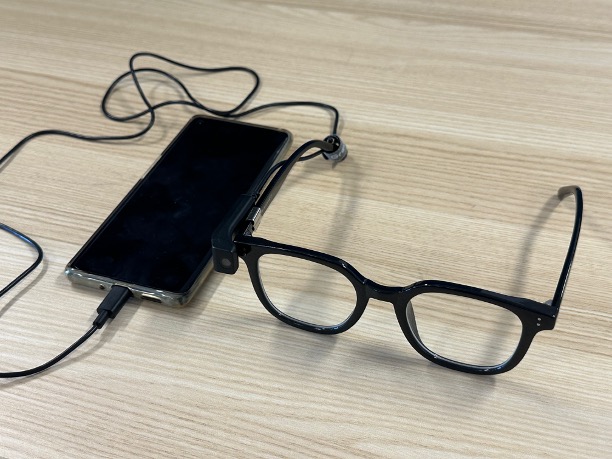}
    \caption{Wearable apparatus for head-aligned first-person audiovisual capture. The glasses-mounted Groudchat camera--microphone module connects via USB Type-C to an Android smartphone that supplies power, stores the recordings, and runs the companion application.}
    \label{fig:set_up}
\end{figure}

% Apparatus 类似aiget即可，api在system design说，具体implementation可以在appendix说
To capture naturally occurring laughter together with head-aligned first-person audiovisual context, participants wore Groudchat\footnote{\url{https://www.groudchat.com/}}, a camera--microphone module mounted on the temple of a pair of glasses.
The module weighs 6.8~g, provides an 88\textdegree{} field of view, and was configured to record 1080p video at 30~fps. It connected to an Android smartphone through USB Type-C. The smartphone supplied power, stored the audiovisual recordings, and ran the companion application. 
The complete setup is shown in Figure~\ref{fig:set_up}.
This hands-free configuration did not require participants to hold or aim a phone during recording and kept the captured audio and video temporally aligned.

% Shared Capture Pipeline
During participant-initiated recording sessions, the application streamed microphone audio to the Speechmatics\footnote{\url{https://www.speechmatics.com/}} real-time audio-event API. The API returned timestamped laughter events with associated information (see Appendix~\ref{sec:appendix-detector-selection} for details). The application retained events above the study-specific confidence threshold, grouped temporally adjacent laughter bouts into candidate moments, and aligned them with the locally stored recording. Each candidate was then associated with bounded audiovisual and spatiotemporal context, producing a laughter-indexed candidate moment for subsequent review. Detector benchmarking, event-construction rules, context-acquisition mechanisms, and parameter settings are reported in Appendix~\ref{app:implementation}.
\section{Formative Study}
% 两页半

Using participants' own laughter episodes, we examined how they interpreted laughter as an index of potentially positive experiences, which contextual cues supported reconstruction of the underlying episode, and how candidate moments should be organized and resurfaced to support reflection. The findings yielded five design considerations that informed the reconstruction and reflection mechanisms of \textit{LaughAnchor}.

\subsection{Participants}

We recruited 12 participants (8 female, 4 male; age: $M = 24.6$, $SD = 3.1$) through a campus forum and social media. Participants reported varied recording and prior-record review practices, as shown in Table~\ref{tab:formative_demographic}.
The study was approved by the Institutional Review Board (IRB) of our university. 
Participants were compensated 10 USD/hour for the interview-based phase and 6 USD per valid session during the recording phase.

\begin{table}[b]
\centering
\begin{tabular}{ccccc}
\toprule
ID & Gender & Age & Recording Practice & Prior-Record Review \\
\midrule
\midrule
P1  & M & 20 & Occasional & Active \\
P2  & F & 23 & Regular   & Occasional   \\
P3  & F & 23 & Regular   & Active \\
P4  & M & 32 & Occasional & Occasional  \\
P5  & F & 24 & Regular   & Active \\
P6  & F & 28 & Regular   & Active \\
P7  & M & 21 & Regular   & Active \\
P8  & F & 25 & Occasional & Occasional  \\
P9  & F & 26 & Regular   & None   \\
P10 & F & 24 & Occasional & Active \\
P11 & M & 25 & Regular   & Occasional   \\
P12 & F & 24 & Occasional & Occasional  \\
\bottomrule
\end{tabular}
\caption{Participant characteristics and self-reported pre-study practices. Recording practice was categorized as \textit{Regular}, \textit{Occasional}, or \textit{None}; prior-record review practice was categorized as \textit{Active}, \textit{Occasional}, or \textit{None}.}
\label{tab:formative_demographic}
\end{table}

\subsection{Study design and procedures}

% 为什么不能只通过普通访谈讨论一个假想 laughter system，而需要先采集参与者自己的自然笑声，再让其进行 think-aloud？

The capture pipeline described in Section~\ref{sec:capture-pipeline} automatically generated candidate laughter episodes. But these episodes were initially raw and fragmented traces, with isolated laughter clips detached from the surrounding narrative and not yet integrated into a meaningful experience. As \citet{staahl2009experiencing} argue, a system should not dictate how such material is to be interpreted, while still enabling the user to participate in the interpretive act. Remembering is reconstructive, with contextual cues helping people rebuild experienced events~\cite{rubin2015event,van2008informing}. Reflection also involves intellectual and affective engagement that can produce new understanding or appreciation~\cite{fleck2010reflecting}. Guided by these principles, we designed the study to keep interpretation user-led and avoid assigning meaning to captured materials in advance. The sequential procedure therefore used participants' own laughter episodes rather than hypothetical examples.

% 概述三个研究阶段，并强调它们之间的数据依赖关系
% As shown in Figure~\ref{fig:formative_study_procedure}, the study comprised three stages. \textbf{1) Semi-Structured Pre-Study Interviews.} We examined participants' recording and prior-record review practices, perceptions of passive laughter capture, acceptable recording boundaries, and expectations for contextual support. \textbf{2) Natural Behavior Recording.} Over seven consecutive days, participants recorded self-selected everyday activities for 1.5--2.5 hours per day. After each session, they briefly described the activity, social setting, and subjective experience. These recordings provided participant-specific materials for subsequent review. \textbf{3) Think-Aloud Review.} Within 1--2 days after the recording period, participants attended a think-aloud session using layered mockups constructed from selected laughter-indexed episodes.

\begin{figure*}[t]
    \centering
    \includegraphics[
        width=\textwidth,
        trim={40mm 60mm 40mm 55mm},
        clip
    ]{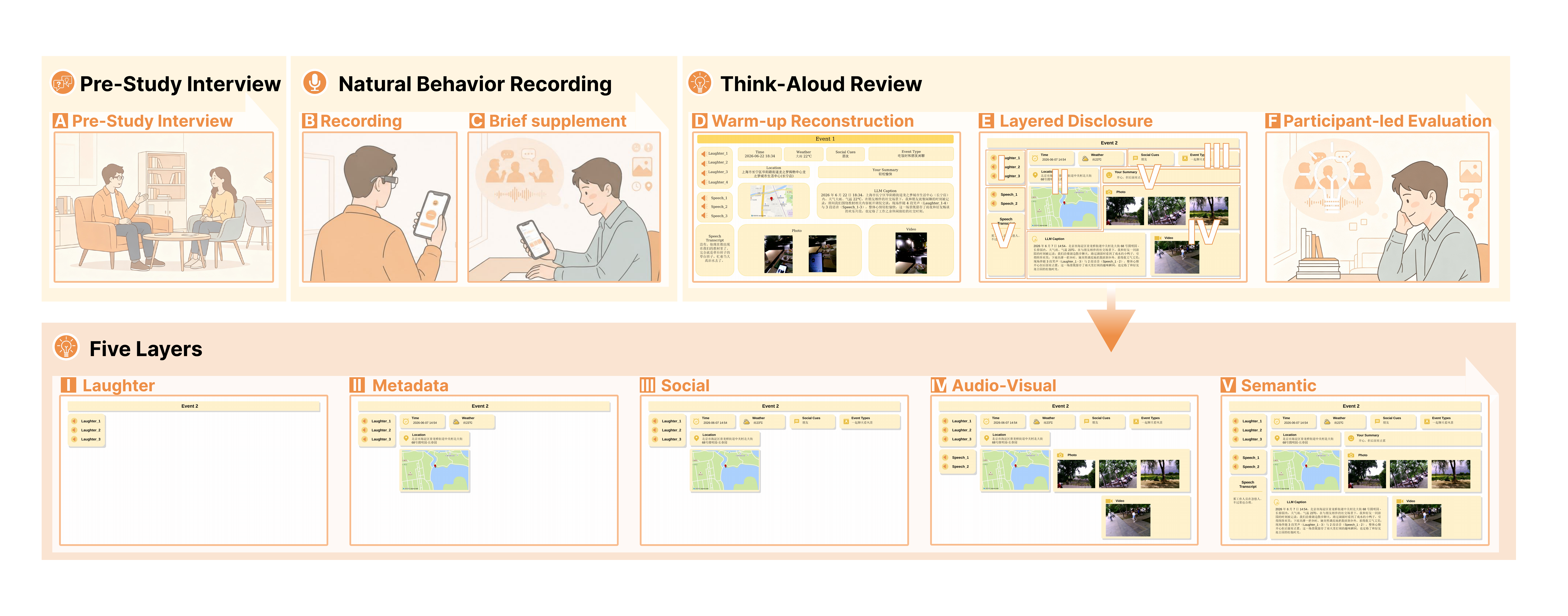}
    \caption{Overview of the formative study procedure and layered think-aloud materials. Participants completed a pre-study interview (A), natural recording with brief post-session supplements (B--C), and a think-aloud review comprising a full-context reconstruction warm-up, cumulative context disclosure, and participant-led evaluation (D--F). The lower panel illustrates the five cumulative context layers used during review, progressing from laughter alone to metadata, social, audiovisual, and semantic context.}
    \label{fig:formative_study_procedure}
\end{figure*}

As shown in Figure~\ref{fig:formative_study_procedure}, participants first completed a pre-study interview covering their recording and prior-record review practices, perceptions of passive laughter capture, acceptable recording boundaries, and expectations for contextual support. They then recorded self-selected everyday activities for 1.5--2.5 hours per day over seven consecutive days and, after each session, briefly described the activity, social setting, and subjective experience. One to two days after the recording period, participants attended a think-aloud session using layered mockups constructed from their own laughter-indexed moments. For each participant, we selected the most recent and earliest moments from the seven-day recordings as near-term and long-term examples, respectively.

% Prior work + interviews determined context types.
% Context was organized into L0–L4 for controlled progressive disclosure.
% Table summarizes what each layer presented and what it probed.

% Think-Aloud Structure的四个任务咋写 要写吗 先写上了
% To construct the think-aloud mockups, we drew on established context frameworks, prior representations of personal laughter, and smart-journal metadata~\cite{abowd1999towards,yang2022exploring,elsden2016s}. We refined the contextual cues through the pre-study interviews and organized them into five cumulatively disclosed layers, progressing from laughter alone to semantic context. Following the distinction between familiarity and richer remembering~\cite{sellen2007life}, we ordered the layers according to their expected contribution to reconstruction and increasing content specificity. This sequence also allowed us to examine privacy concerns~\cite{bellotti1993design} and perceived cognitive burden. Table~\ref{tab:context-scaffold} summarizes the source and purpose of the information introduced at each layer.

Primary context types, including location, identity, time, and activity~\cite{abowd1999towards}, and prior temporal, spatial, and social representations of personal laughter~\cite{yang2022exploring} informed our selection of metadata and social cues. We also included first-person audiovisual materials to probe perceptual and interactional detail, and semantic annotations informed by user-authored entries and tags in smart journals~\cite{elsden2016s}. We refined this cue set through the pre-study interviews and organized the mockups into five cumulative layers, beginning with laughter alone.
Drawing on the distinction between familiarity and richer remembering~\cite{sellen2007life}, we ordered the layers by increasing content specificity to examine how additional context supported reconstruction. This sequence also allowed us to examine privacy concerns~\cite{bellotti1993design} and perceived cognitive burden. Table~\ref{tab:context} summarizes the source and purpose of the information introduced at each layer. Implementation details of the formative capture configuration are reported in Appendix~\ref{app:formative-context-configuration}.

The think-aloud review began with a brief warm-up in which participants viewed a separate example from their own recordings and shared their initial impressions. We then progressively disclosed the five context layers for each of the two selected moments, one near-term and one long-term.
At each layer, they described what they recognized, which new details emerged, and whether the information was sufficient, redundant, private, or burdensome. Follow-up probes examined affective re-experiencing, new understanding or appreciation, and preferences for retaining, deleting, or resurfacing moments across different times and situations. The study focused on participant-led reconstruction and reflection, but not objective memory accuracy.

\begin{table*}[b]
\centering
\begin{tabularx}{\textwidth}{
@{}
>{\raggedright\arraybackslash}p{0.14\textwidth}
>{\raggedright\arraybackslash}p{0.25\textwidth}
>{\raggedright\arraybackslash}p{0.16\textwidth}
>{\raggedright\arraybackslash}X
@{}
}
\toprule
\textbf{Layer} &
\textbf{New Information Added} &
\textbf{Source} &
\textbf{Purpose in the Think-Aloud} \\
\midrule
\midrule
L0: Laughter &
Laughter audio &
Automatically captured &
Probe whether laughter alone supported familiarity, affective access, or episode identification. \\
\hline
L1: Metadata &
Time, location, and weather &
Automatically captured &
Probe whether spatiotemporal metadata helped participants orient the laughter cue to an everyday episode. \\
\hline
L2: Social &
Activity type, social setting, and people present &
Participant-reported after each recording session &
Probe whether activity and social information disambiguated the event and its interactional setting. \\
\hline
L3: Audio-Visual &
First-person photos, short video, and surrounding audio &
Automatically captured &
Probe whether perceptual and interactional cues supported richer reconstruction of the scene and laughter trigger. \\
\hline
L4: Semantic &
User-authored summary, transcript, and LLM-generated caption &
Participant-authored or system-generated &
Probe semantic confirmation, redundancy, and the risk of constraining or replacing user interpretation. \\
\bottomrule
\end{tabularx}
\caption{Context layers used in the formative think-aloud study. Layers were disclosed cumulatively from L0 to L4, with each layer adding the information shown in the second column.}
\label{tab:context}
\end{table*}

\subsection{Data analysis}

We analyzed the pre-study interviews and think-aloud sessions using qualitative thematic analysis. Two authors independently coded 60\% of the data and discussed their codes to develop preliminary themes. The research team coded the remaining 40\% with reference to this emerging structure and iteratively refined the thematic map as new patterns and relationships emerged. This process yielded four themes: \textit{Affective Trace}, \textit{Moment-Dependent Scaffold}, \textit{Meaning Beyond Capture}, and \textit{Situated Reflection}. The first two themes explain how contextual grounding transforms detected laughter into a reconstructable episode, while the latter two concern how captured episodes acquire personal value and when they should reappear.

Moreover, to complement the thematic analysis, we assessed the richness of participants' verbal reconstructions using an adapted rubric informed by the \textit{Autobiographical Interview}~\cite{levine2002aging}. Prior self-tracking systems have also adapted autobiographical recall measures for their evaluation objectives~\cite{sas2013affectcam,niforatos2015pulsecam}. Specifically, we rated reconstruction richness across five dimensions covering event, place, time, perceptual information, and thought/emotion, each on a 0--3 scale. At each disclosure layer, the five ratings were summed to yield a cumulative score of 0--15. We used the total and dimension-level scores to describe reconstruction richness and stage-wise changes within the cumulative disclosure sequence. Detailed scoring criteria and rating procedures are provided in Appendix~\ref{app:scoring-procedures}.

\subsection{Findings}

\subsubsection{F1. Laughter serves as an affective index, while corresponding context supports reconstruction of the lived experience.}

% laughter的情绪效力与感染力
Participants described laughter as a dynamic, relatively unprocessed affective trace that reinstated the emotional tone, atmosphere, and interactional qualities of an episode. Its rhythm, tone, and surrounding audio carried \textit{"dynamic information"} (P4) and made the experience \textit{"feel vivid and affectively contagious"} (P3). P7 perceived subtle differences across laughter instances, indicating that audio preserved subjective nuances that static photos or text conveyed less readily.
% 笑声本身无法定位准确事件
However, laughter alone did not reliably identify a specific episode. Participants sometimes recognized a clip as positive or familiar without recalling what had happened or why they had laughed, especially when the underlying episode was temporally distant, offered few distinctive semantic cues, or occurred within a repetitive activity. P3, for example, recovered only the \textit{"overall rhythm"} of an extended enjoyable game because \textit{"everyone was happy throughout the game"} and no distinctive cue linked the laughter to a particular interaction. Presumably, laughter supported coarse affective recognition but was insufficient for reliable episodic reconstruction.

% 需要笑声结合附近的context info,尽管不能直接定位，但是laughter作为情绪入口仍能够吸引/鼓励人去resurface
Reconstruction emerged when laughter aligned with episode boundaries and proximal context. Time, location, and social information first situated the episode, while nearby speech and audiovisual cues then recovered dialogue, interactions, and the laughter trigger. As P10 explained, understanding required knowing \textit{"what happened right before and after this laughter"}. These accounts suggest that a minimally reconstructable unit combined laughter with temporally and semantically aligned context. The cues required varied with temporal distance, episode distinctiveness, scene repetition, and whether the laughter came from social interaction or solitary enjoyment.
Notably, the value of laughter also extended beyond episodic identification. P2 spontaneously asked, \textit{"What was I laughing so foolishly about back then?!"} This response illustrated how the affective force of laughter could invite further exploration.

\subsubsection{F2. Temporal distance influences both reconstruction challenges and contextual needs for resurfacing positive moments.}

% long/short的reconstruct的 需求/challenge 差异
% long的特征
% short的特征
Temporal distance changed what participants sought to reconstruct and how laughter functioned within a laughter-indexed moment. In long-term reviews, event details had often faded while a general affective impression remained. Participants were often more interested in re-entering the positive feeling than recovering every detail. Fortunately, laughter could still reinstate this feeling despite weaker episodic grounding, serving as a residual affective trace, while time and location offered low-cost anchors for locating the episode.
In near-term reviews, participants generally recognized the activity but needed to recover the specific interaction. When an episode was distinctive, laughter often reactivated it with little additional context. When several recent experiences were similar, laughter provided an initial direction for reconstruction, and photos, video, and nearby audio helped disambiguate the interaction and laughter trigger. Therefore, long-term reviews often began with affective re-entry followed by episode grounding, while near-term reviews focused more on disambiguation and detail recovery.

% Descriptively, the reconstruction-richness profiles supported this distinction (Figures~\ref{fig:richness_trajectory} and~\ref{fig:richness_heatmap}). For earlier episodes, the largest Emotion gain occurred at L0 laughter and L3 reactivated more Perceptual and Event details. For recent episodes, reconstruction began from a higher baseline and gains were distributed more broadly across dimensions. Within the Event dimension, L0 laughter and L3 audio-visual context made the largest and second-largest contributions. 
% Across both conditions, gains diminished once participants had reconstructed a coherent scene, particularly at L4. Participants described a cascade-like process in which later layers mainly provided \textit{"secondary verification"} (P12) or made the information \textit{"more cluttered"} (P5).
% It indicates that temporal distance shaped both the reconstruction target and the role of laughter within resurfacing. Earlier moments were more oriented toward re-entering an emotionally consolidated experience, and laughter served as a residual affective trace here. Recent moments were more oriented toward identifying and recovering a specific recent episode, where laughter usually served as an initial index for identifying the event.

\begin{figure*}[t]
    \centering
    \begin{subfigure}[t]{0.40\textwidth}
        \centering
        \includegraphics[width=\linewidth]{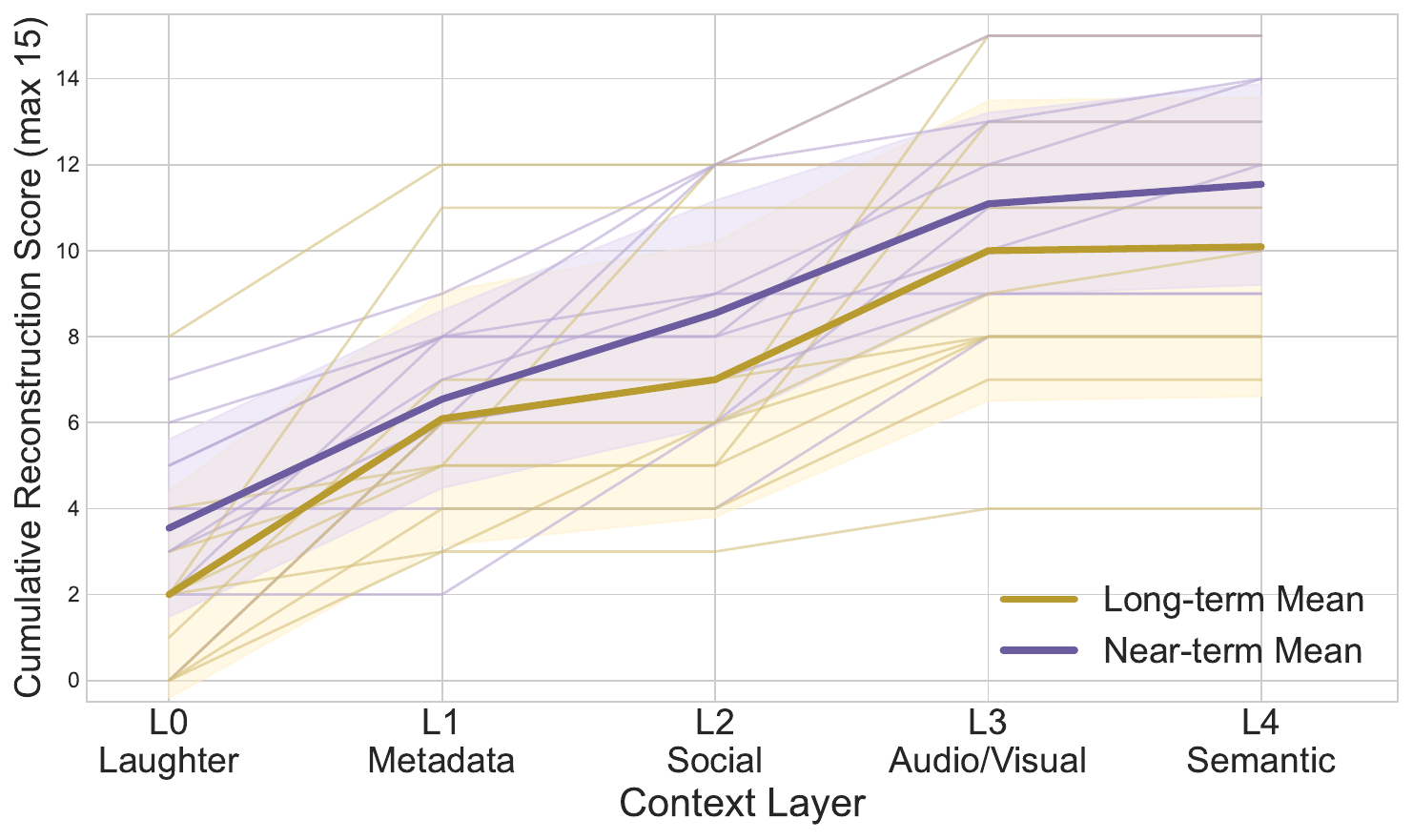}
        \caption{Cumulative reconstruction richness across context layers.}
        \label{fig:richness_trajectory}
    \end{subfigure}
    \hfill
    \begin{subfigure}[t]{0.57\textwidth}
        \centering
        \includegraphics[width=\linewidth]{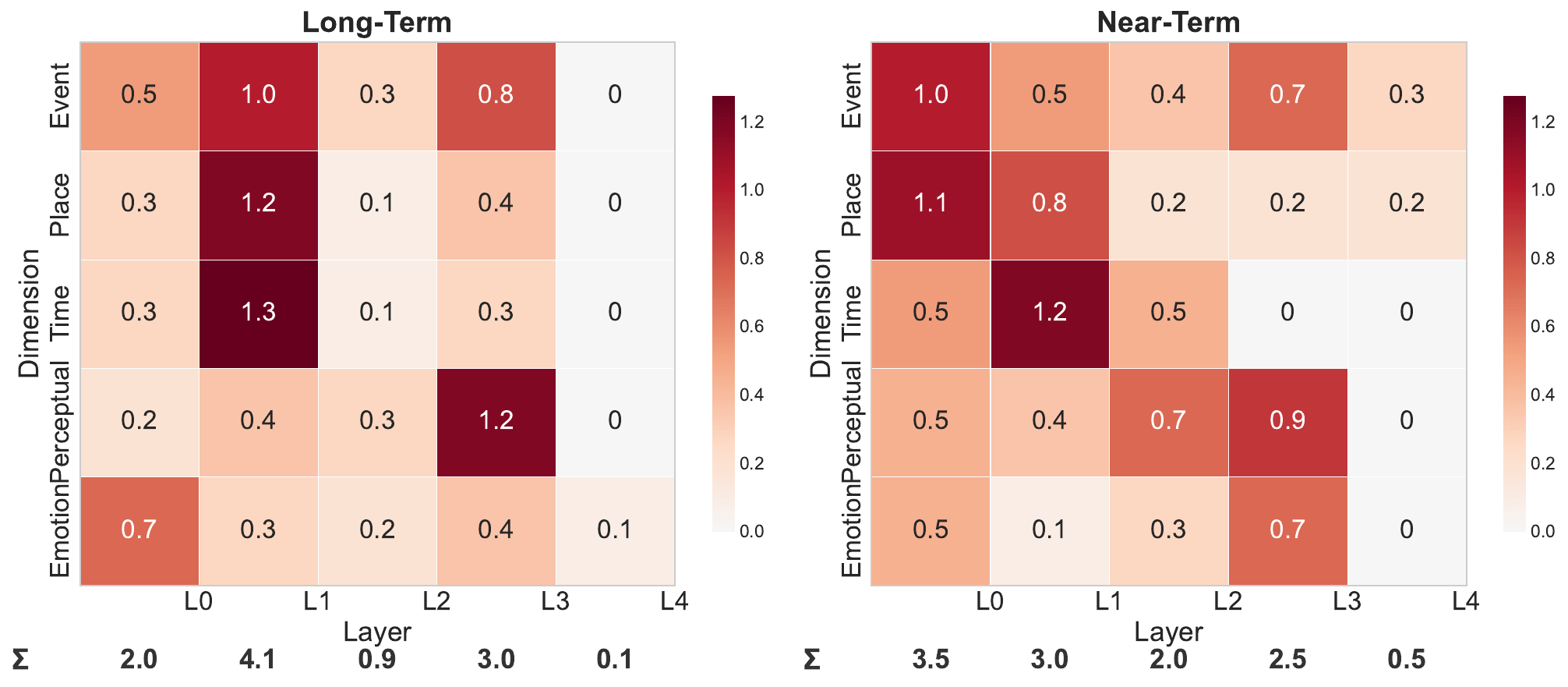}
        \caption{Marginal contributions by context layer and detail dimension.}
        \label{fig:richness_heatmap}
    \end{subfigure}
    \caption{Reconstruction richness during cumulative context disclosure for near-term and long-term episodes. The trajectory plot shows cumulative richness scores (0--15) across L0--L4, while the heatmaps show mean marginal gains by internal-detail dimension and context layer.}
    \label{fig:richness_analysis}
\end{figure*}

Descriptively, the reconstruction-richness profiles supported this distinction (Figures~\ref{fig:richness_trajectory} and~\ref{fig:richness_heatmap}). For long-term reviews, the largest gain in Emotion occurred at L0 laughter, while L3 recovered more Perceptual and Event details. For near-term reviews, reconstruction began from a higher baseline, and gains were distributed more broadly across dimensions. Within the Event dimension, L0 laughter and L3 audiovisual context made the largest and second-largest contributions. 
Across both conditions, gains diminished once participants had reconstructed a coherent scene, especially at L4. Participants described a cascade-like process in which later layers mainly offered \textit{"secondary verification"} (P12) or made the information \textit{"more cluttered"} (P5). 
% These profiles support the qualitative distinction between a residual affective role for laughter in long-term reviews and an indexing role in near-term episode identification.

\subsubsection{F3. Passive capture preserved candidate traces of overlooked moments, and reflective value emerged through user interpretation.}

% passive capture捕捉未注意到细节

Participants valued passive laughter capture because it reduced in-the-moment recording effort and preserved brief positive experiences that they might otherwise not have documented. P7 enthusiastically compared the later discovery of these clips to \textit{"treasure hunting",} while P3 described the records as \textit{"very real and natural".} Automatic capture also accumulated ordinary, repetitive, or otherwise low-value moments. The value of passive capture therefore lay in broadening the set of candidate moments available for later consideration and bringing overlooked experiences into users' awareness with little in-the-moment effort.

% 需要使用者engagement来决定保留价值
Notwithstanding these advantages, participants distinguished preserving the occurrence of laughter from assigning meaning to it. Although laughter could signal a potentially positive episode, it did not establish reflective value or determine whether a moment should be retained or resurfaced. These judgments depended on relationships, authenticity, emotional salience, self-relevance, event distinctiveness, and specific content (P6, P7, P10). Participants did not view retrospective value as binary. P7 noted that it could \textit{"vary with context",} making fixed automated criteria inappropriate.
%% ownership ; 对ai-generated的反感 interpret / engagement的时机可以是事后的
Participants also wanted authority over how and when meaning was constructed. P10 felt that one LLM-generated summary had erased the record's \textit{"original meaning".} As P12 noted, describing a memory \textit{"inevitably carries the speaker's own view and understanding of the experience"}. 
Automatically generated accounts could feel as if they replaced personal recollection instead of supporting it. Immediate interpretation was also often impractical because participants were \textit{"immersed in the moment or busy doing something else".} 
% They preferred to interpret and supplement moments later, at a more suitable time. Taken together, passive capture preserved candidate traces, whereas reflective value emerged through subsequent user interpretation and decisions about retention and future resurfacing.
They preferred to interpret and supplement captured moments later, when they could assess their significance and decide whether to retain or resurface them.

% \subsubsection{F4. Positive moment reflection provides distinct benefits across temporal distances.}

\subsubsection{F4. Near-term and long-term resurfacing support complementary forms of reflection.}

% long/short 去resurface不同的好处
Participants described different benefits at the two temporal distances. In near-term reviews, affective residue from the original experience often remained, allowing resurfacing to reinforce or clarify an ongoing positive feeling. P3 compared positive affect to a gradually declining peak and noted that resurfacing a moment while some emotion remained made the response more salient. Participants also described that fatigue or tension associated with the original event could fade while enjoyment remained, producing a more positive recollection consistent with the \textit{rosy-view} effect~\cite{mitchell1997temporal}.
Long-term resurfacing helped participants rediscover ordinary moments as part of broader patterns in their relationships and everyday lives. P2 felt that resurfacing made life seem \textit{"richer and more substantial"}, while P12 became more aware of \textit{"many small moments of warmth"}. Participants did not describe fundamental changes in their overall evaluation of life. Resurfacing reinforced, elaborated, or made visible positive experiences that everyday busyness had obscured. Thus, near-term resurfacing primarily supported affective re-experiencing, while long-term resurfacing supported rediscovery and meaning consolidation. These benefits were complementary, and the meaning of the same moment could continue to develop through repeated resurfacing at different temporal distances.

% \subsubsection{F5. Reflection should preserve user agency while providing low-interruption support.}
\subsubsection{F5. Resurfacing should remain optional, low-interruption, and under user control.}

% 不希望被强制要求回顾
% 也不希望完全靠自己回顾
Participants rejected forced, frequent, or attention-demanding resurfacing. They preferred access to remain primarily self-directed, with occasional reminders presented as optional and dismissible invitations. P7 described such a reminder as a \textit{"pleasant surprise"}, while P2 emphasized that \textit{"the less it interrupts me, the better"}. Forced viewing could itself \textit{"feel disruptive"} (P12).
At the same time, participants did not want resurfacing to depend entirely on deliberate retrieval. P9 noted that they rarely sought out past records, but an occasional prompt could lead them to engage with a moment they would otherwise overlook. Participants therefore favored a balance between user initiative and lightweight system support. Situated cues offered one way to achieve this balance. P10 suggested resurfacing a past moment when the user returned to the same location. More broadly, participants considered resurfacing most appropriate when it fit naturally into everyday activity, remained easy to ignore, and did not require deliberate retrieval.

\begin{table*}[t]
\centering
\small
\renewcommand{\arraystretch}{1.22}
\begin{tabularx}{\textwidth}{
@{}
>{\raggedright\arraybackslash}p{0.19\textwidth}
>{\raggedright\arraybackslash}p{0.40\textwidth}
>{\raggedright\arraybackslash}X
@{}
}
\toprule
\textbf{Theme} &
\textbf{Empirical Finding} &
\textbf{Design Consideration} \\
\midrule
\midrule
\textbf{Affective Trace}
&
\multirow[t]{2}{=}{%
\textbf{F1.} Laughter serves as an affective index, while corresponding context supports reconstruction of the lived experience.
}
&
\multirow[t]{2}{=}{%
\textbf{DC1.} Progressively ground laughter-indexed affective traces in reconstructable episodes through layered context.
}
\\
\cmidrule{1-1}

\multirow[t]{2}{=}{%
\textbf{Moment-Dependent Scaffold}
}
&
&
\\
\cmidrule{2-3}

&
\textbf{F2.} Temporal distance influences both reconstruction challenges and contextual needs for resurfacing positive moments.
&
\textbf{DC2.} Adapt contextual presentation to temporal distance and resurfacing demands, accounting for the different roles of laughter cues in reconstruction.
\\
\midrule

\textbf{Meaning Beyond Capture}
&
\textbf{F3.} Passive capture preserved candidate traces of overlooked moments, and reflective value emerged through user interpretation.
&
\textbf{DC3.} Use passive laughter sensing to surface candidate moments while preserving user authority over interpretation, selection, and retention.

\\
\midrule

\multirow[t]{2}{=}{%
\textbf{Situated Reflection}
}
&
\textbf{F4.} Near-term and long-term resurfacing support complementary forms of reflection.
&
\textbf{DC4.} Support complementary near-term and long-term resurfacing of laughter-indexed moments.
\\
\cmidrule(lr){2-3}

&
\textbf{F5.} Resurfacing should remain optional, low-interruption, and under user control.
&
\textbf{DC5.} Keep resurfacing optional, low-interruption, and sensitive to the current context.
\\

\bottomrule
\end{tabularx}

\caption{Mapping of the four formative-study themes to five empirical findings and five design considerations. The staggered multirow structure indicates that F1 and DC1 span both \textit{Affective Trace} and \textit{Moment-Dependent Scaffold}.}
\label{tab:formative-analysis}
\end{table*}
\section{\textit{LaughAnchor} System Design}

Based on the formative findings summarized in Table~\ref{tab:formative-analysis}, we derived five design considerations (DCs) that guided the capabilities and interaction design of \textit{LaughAnchor}, as following:

\paragraph{\textbf{DC1. Progressively ground laughter-indexed affective traces in reconstructable episodes through layered context.}}

Context should support a progression from laughter-based affective recognition to episode localization and richer reconstruction through temporally aligned cues when needed.

\paragraph{\textbf{DC2. Adapt contextual presentation to temporal distance and resurfacing demands, accounting for the different roles of laughter cues in reconstruction.}}

For long-term resurfacing, prioritize spatiotemporal anchors and reveal richer context on demand. For near-term resurfacing, foreground cues that help users distinguish similar episodes and recover specific interactions and laughter triggers.

\paragraph{\textbf{DC3. Use passive laughter sensing to surface candidate moments while preserving user authority over interpretation, selection, and retention.}}

Automatic capture should broaden the candidate set without assigning meaning or value. Users should be able to supplement or delete candidates, retain them with or without proactive resurfacing, and revise these decisions asynchronously.

\paragraph{\textbf{DC4. Support complementary near-term and long-term resurfacing of laughter-indexed moments.}}

Near-term resurfacing should support affective re-experiencing, whereas long-term resurfacing should support rediscovery and meaning consolidation. The system should not assume a single optimal interval.

\paragraph{\textbf{DC5. Keep resurfacing optional, low-interruption, and sensitive to the current context.}}

Prompts should remain dismissible and non-demanding, while optional situated cues may provide occasional resurfacing opportunities when appropriate.

\subsection{Key Elements of \textit{LaughAnchor}}

\begin{figure}
    \centering
    \includegraphics[width=\linewidth]{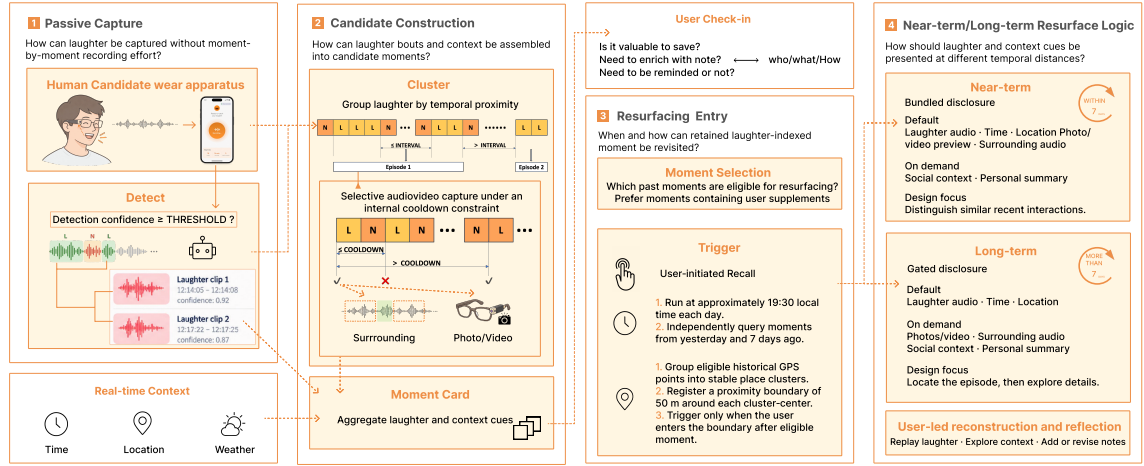}
    \caption{Overview of the \textit{LaughAnchor} processing pipeline. Passive capture and context collection (1) provide laughter and contextual cues for candidate Moment Card construction (2). User check-in supports supplementation and decisions about retention and future resurfacing. Retained moments can be revisited through user-initiated access or time- and place-based suggestions (3). Bundled disclosure for near-term resurfacing and gated disclosure for long-term resurfacing (4) provide different starting cues for user-led reconstruction and reflection.}
    \label{fig:system-dataflow}
\end{figure}

We implemented the five DCs through three connected design elements centered on the Moment Card, which binds detected laughter to contextual materials from the surrounding episode.
Figure~\ref{fig:system-dataflow} summarizes this workflow, while Figure~\ref{fig:app-workflow} illustrates the corresponding interactions. The following sections explain how these implement DC1--DC5.

\subsubsection{Layered context disclosure (\textbf{DC1, DC2}).}

Guided by \textbf{DC1}, we designed each Moment Card with default and expandable context to support a progression from laughter-based affective recognition to reconstruction of the underlying episode. To keep the recall process lightweight, the interface initially presents selected cues; users can stop when these are sufficient or choose to reveal further context as needed.

To apply \textbf{DC2}, we adapted the amount and ordering of initially visible context to temporal distance (Figure~\ref{fig:app-workflow}b). For long-term reviews, the system uses \textbf{gated disclosure}, with playable laughter audio, capture time, and location in the default view. Laughter provides an affective entry point, while the spatiotemporal cues help users locate the episode before consulting more detailed material. Photos, video, surrounding audio, social context, and the user-authored summary remain expandable.

For near-term reviews, the system uses \textbf{bundled disclosure}. The default view combines playable laughter audio, a photograph or video preview, capture time, location, and surrounding audio. Presenting these cues together is intended to help users distinguish similar recent interactions and recover the specific conversation or laughter trigger with fewer expansion steps. Social context and the user-authored summary remain expandable.

\subsubsection{Passive capture, automated assembly, and user-curated meaning-making (\textbf{DC3}).}

To implement \textbf{DC3}, \textit{LaughAnchor} automates capture and initial organization while leaving their interpretation and preservation to the user. During user-initiated recording periods, the shared pipeline (described in Section~\ref{sec:capture-pipeline}) groups detected laughter bouts into candidate moments and associates each with bounded contextual cues. 
User can review the resulting Moment Cards later, without having to decide as an experience unfolds whether it is worth preserving.

Candidate moments are retained by default unless users select \textit{`Delete'}. The options \textit{`Save, allow resurfacing suggestions'} and \textit{`Save, but don't resurface this'} both preserve self-directed access, but only the former makes a moment eligible for future system suggestions.

User control also extends beyond the initial check-in (\textbf{DC3}). Users can revise retention and resurfacing decisions after subsequent reviews, and add or edit personal summaries and comments at any time using text, voice, or uploaded photos. This asynchronous, revisable workflow allows users to develop their own accounts when they have time to revisit the experience.

\subsubsection{Complementary, low-interruption resurfacing (\textbf{DC4, DC5}).}

In support of \textbf{DC4}, \textit{LaughAnchor} provides resurfacing opportunities at both near-term and long-term intervals. Once users have reviewed a candidate and authorized future suggestions, the same retained moment can return at different temporal distances. This design accommodates both revisiting a still-accessible positive experience and rediscovering it later, with bundled or gated disclosure shaping how each review begins.

To minimize interruption (\textbf{DC5}), the system delivers resurfacing suggestions as dismissible phone notifications that do not require an immediate response. Users may ignore a suggestion, open the Moment Card when convenient, or subsequently change whether the moment remains eligible for proactive resurfacing.

The contextual aspect of \textbf{DC5} also informs optional \textbf{situated resurfacing}. When users return near the capture location of a moment authorized for resurfacing, the system may suggest it. Returning to the location provides an opportunity for resurfacing, while users decide whether the current situation is suitable for review.

\subsubsection{Archive-level visualization of laughter-indexed moments (\textbf{supplementary design}).}

\begin{figure}
    \centering
    \includegraphics[
        width=\linewidth,
        trim={5mm 5mm 6mm 5mm},
        clip
    ]{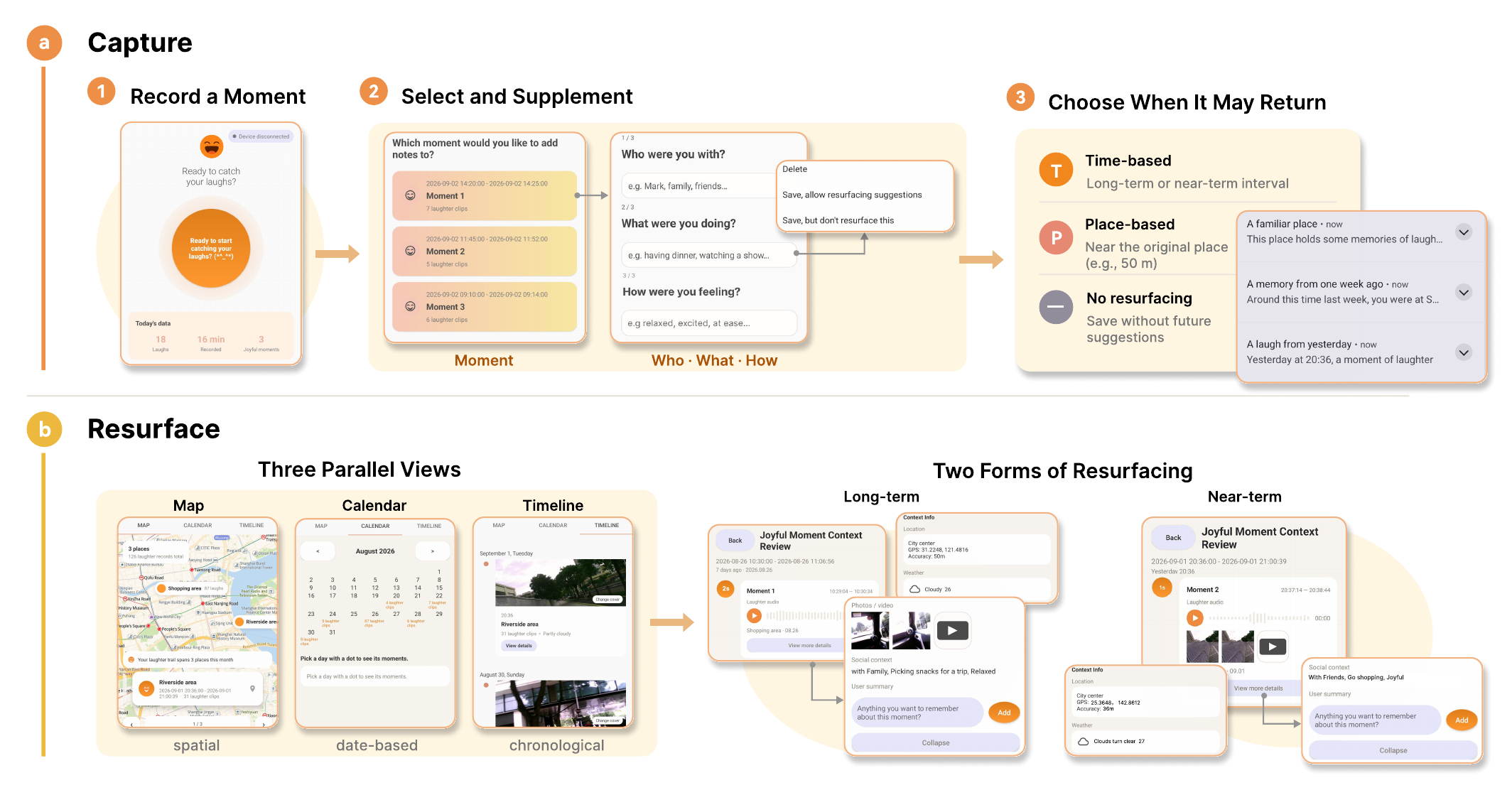}
    \caption{Core interactions in \textit{LaughAnchor}: (a) recording, supplementation and curation of candidate Moment Cards, and decisions about time- or place-based resurfacing; (b) self-initiated browsing through Map, Calendar, and Timeline views, with bundled disclosure for near-term reviews and gated disclosure for long-term reviews.}
    \label{fig:app-workflow}
\end{figure}

Beyond the individual Moment Cards used in the formative study, \textit{LaughAnchor} provides three parallel views for temporal and spatial exploration (Figure~\ref{fig:app-workflow}b). The \textbf{Calendar View} supports date-based access, the \textbf{Timeline View} organizes moments chronologically, and the \textbf{Map View} provides location-based cues for exploring where moments occurred. All three views support self-initiated access to retained moments, including those without proactive resurfacing.

The interface also displays descriptive summaries of laughter frequency and duration. Together with the temporal and spatial views, these summaries are intended to provide a lightweight entry point for observing patterns in recorded experiences and reflecting on everyday routines and social rhythms. We cautiously present these quantities without treating more frequent or longer laughter as a better outcome, leaving their personal significance for users to interpret.

\subsection{Apparatus and Implementation}

The field prototype retained the shared sensing and laughter-indexing architecture described in Section 3. The detector was selected based on benchmarking with public datasets (Appendix~\ref{sec:appendix-detector-selection}), while the formative study further demonstrated the practical feasibility of the overall pipeline by detecting laughter and generating laughter-indexed moments from participants' everyday recordings. Building on these findings, the primary iteration focused on the interface and display logic of the companion app, implementing the design considerations derived from the formative study. We also adjusted several system parameters according to observed performance and participants' feedback. Further implementation details, including candidate construction and clustering, context acquisition, resurfacing scheduling, and study-specific parameter settings, are reported in Appendix~\ref{app:implementation}.

\section{Field Study}
%2页半

We conducted a three-week in-the-wild deployment of \textit{LaughAnchor} with 12 participants to examine how they used and interpreted laughter-indexed moments in everyday life. The study examined the following four aspects of use, which collectively address RQ1 and RQ2:

\textbf{S1. Coverage:} Passive laughter indexing complemented existing recording practices by surfacing participant-valued moments that would otherwise be unlikely to be recorded. 

\textbf{S2. Affective Reconnection:} Laughter-indexed resurfacing provided a distinct affective route back to the emotion, atmosphere, and interactional qualities of an experience. 

\textbf{S3. Reconstructive and Reflective Value:} Context-scaffolded near-term and long-term resurfacing of laughter-indexed moments supported episode reconstruction and complementary forms of reflection, including rediscovery awareness of recent life, relationships, and emotional states. Across moments and reviews, these outcomes further contributed to meaning consolidation.

% \textbf{S4. Exploratory Well-being:} The rhythm and depth of system use may be directionally associated with changes in self-reported well-being.

\textbf{S4. Exploratory Well-being:} Self-reported well-being scores increased overall during deployment, while participants' explanations of perceived emotional changes included both system use and everyday circumstances.

\subsection{Participants and Study Design}

Prior works have characterized self-tracking as a lived practice that is embedded in daily routines and varies across users, goals, and stages of engagement~\cite{rooksby2014personal,epstein2015lived,rapp2016personal}. 
% We therefore used participants' existing recording and review practices as participant-specific reference points to evaluate \textit{LaughAnchor} and examine how it complemented these practices.
We therefore used participants' existing recording and review practices as participant-specific reference points for interpreting their experiences with \textit{LaughAnchor} and examining how it complemented these practices.
We did not impose a standardized journaling comparator, and participants were free to continue, suspend, or modify their existing practices during deployment.

\subsubsection{Participants}

We recruited participants through an on-campus communication platform, Xiaohongshu\footnote{\url{https://www.xiaohongshu.com/}}, and participant referrals. We enrolled 12 participants (Male: 2, Female: 10; age: $M = 23.8$, $SD = 4.2$) who anticipated recurring opportunities for spontaneous laughter during everyday conversation, social interaction, leisure, or other activities. Table~\ref{tab:user-study-participants} summarizes their characteristics and prior recording and review practices. All participants completed the study. The study received the ethics approval by the Institutional Review Board of our university, and all participants provided informed consent before attending the study. Participants received a fixed honorarium of USD 45 for completing the study.

\begin{table*}[t]
\centering
\small
\setlength{\tabcolsep}{3.2pt}
\renewcommand{\arraystretch}{1.10}

\begin{tabular*}{\textwidth}{
@{\extracolsep{\fill}}
l
c
c
c
c
c
c
c
c
c
c
c
@{}
}
\toprule

\multirow{2}{*}{\textbf{ID}} &
\multirow{2}{*}{\textbf{Gender}} &
\multirow{2}{*}{\shortstack{\textbf{Age}\\\textbf{(years old)}}} &
\multicolumn{2}{c}{\textbf{Self-Reported Frequency}} &
\multirow{2}{*}{\shortstack{\\\textbf{Perceived}\\\textbf{Omission}}} &
\multicolumn{6}{c}{\textbf{Usual Recording Media}} \\

\cmidrule(lr){4-5}
\cmidrule(lr){7-12}

&
&
&
\shortstack{\textbf{Recording}\\\textbf{(\#/1 week)}} &
\shortstack{\textbf{Review}\\\textbf{(\#/1 month)}} &
&
\textbf{Photo} &
\textbf{Video} &
\shortstack{\textbf{Social}\\\textbf{media}} &
\shortstack{\textbf{Chat}\\\textbf{records}} &
\shortstack{\textbf{Text}\\\textbf{notes}} &
\shortstack{\textbf{Other}\\\textbf{app}} \\

\midrule
\midrule

P1  & F & 24 & 1--2 & 4--8  & Often &
\cmark & \cmark & --           & \cmark & --           & -- \\

P2  & M & 21 & 3--5 & 1--2  & Often &
\cmark & --           & --           & --           & --           & -- \\

P3  & F & 23 & 3--5 & 1--2  & Sometimes &
\cmark & \cmark & \cmark & \cmark & --           & -- \\

P4  & F & 23 & 1--2 & 4--8  & Often &
\cmark & \cmark & \cmark & \cmark & \cmark & -- \\

P5  & F & 25 & 3--5 & 1--2  & Often &
\cmark & \cmark & --           & \cmark & \cmark & \cmark \\

P6  & F & 22 & $\geq 6$ & 1--2  & Often &
\cmark & \cmark & --           & \cmark & \cmark & -- \\

P7  & F & 21 & 3--5 & 1--2  & Sometimes &
\cmark & \cmark & --           & \cmark & --           & -- \\

P8  & F & 21 & 1--2 & 4--8  & Sometimes &
\cmark & \cmark & \cmark & \cmark & --           & \cmark \\

P9  & M & 20 & $\geq 6$ & 4--8  & Often &
\cmark & \cmark & \cmark & \cmark & --           & \cmark \\

P10 & F & 36 & 1--2   & 1--2  & Sometimes &
\cmark & \cmark & --           & --           & --           & \cmark \\

P11 & F & 25 & 3--5 & 1--2  & Often &
\cmark & \cmark & \cmark & --           & --           & -- \\

P12 & F & 24 & 1--2   & 4--8  & Often &
\cmark & --           & \cmark & --           & --           & \cmark \\

\bottomrule
\end{tabular*}
\caption{Participant characteristics and self-reported pre-study recording and review practices. Recording frequency is reported per week and review frequency per month. Perceived omission indicates how often participants reported leaving positive everyday moments unrecorded. Recording-media categories are non-exclusive. \textit{Text notes} includes diaries, journals, and memo applications.}
\label{tab:user-study-participants}
\end{table*}

\subsubsection{Procedure}

% user study流程 (画个timeline流程图)

\begin{figure}[t]
    \centering
    \includegraphics[width=\linewidth]{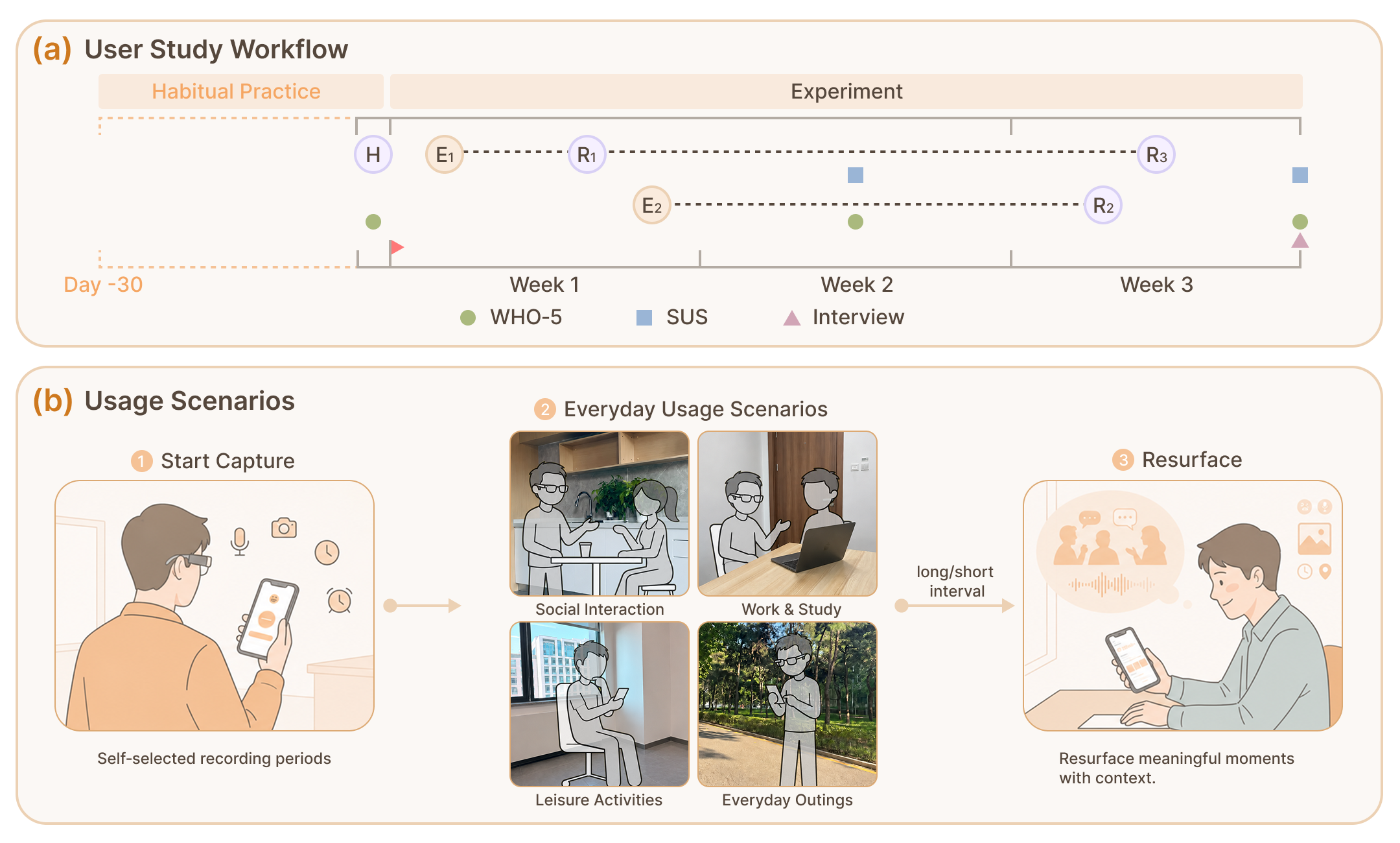}
    \caption{Overview of the field study. (a) A pre-deployment habitual-practice assessment (H) and a three-week deployment with captured events (E) and resurfacing occasions (R). Matching purple markers at H and R indicate the corresponding pre-study habitual and post-resurfacing questionnaires. Dashed lines link example events to subsequent resurfacing occasions, including repeated resurfacing of the same event. (b) Participants initiate recording during self-selected periods across everyday settings and later revisit laughter-indexed moments with contextual cues through near-term and long-term resurfacing.}
    \label{fig:user-study-procedure}
\end{figure}

The study comprised an onboarding session, a flexible in-the-wild deployment, and a post-study session, as shown in Figure~\ref{fig:user-study-procedure}. 
Participants completed the pre-study measures before deployment. During onboarding, they received the apparatus and learned how to initiate recording, inspect and curate candidate moments, and interact with retained moments during resurfacing.
They were also briefed on the system's recording and data-processing workflow and on privacy considerations for themselves and others who might be recorded.
Participants were asked to complete the deployment within three weeks. To accommodate different routines, we imposed no fixed daily schedule or minimum session duration. The recommended minimum use was six recording sessions, four near-term reviews, and three long-term reviews. Participants chose when and where to complete these activities and could undertake additional sessions and reviews at their discretion.
Near-term reviews occurred less than seven days after capture, while long-term reviews occurred seven days or more after capture. The same retained moment could be reviewed at both temporal distances. Participants completed event-level questionnaires following near-term and long-term reviews. The study concluded with post-study measures and a semi-structured interview.
% We intended to give participants experience with the core capture and resurfacing workflows while accommodating their routines and preserving their discretion with this flexible procedure.

\subsubsection{Privacy and Ethical Considerations.}
Participants were instructed to inform anyone who might be recorded, seek their consent, and stop recording if they objected. Recording was participant-initiated, and users could delete candidate moments or retain them without proactive resurfacing. Audiovisual recordings were stored on the smartphone. External processing was limited to streaming microphone audio to Speechmatics for laughter detection.

\subsection{Data Collection}

\subsubsection{Questionnaire}
% 说明 问卷使用、 log中哪些数据的分析

\paragraph{Pre-Study Habitual-Practice Questionnaire.}

Before deployment, participants consulted personal records from the preceding month and reported recording and review frequency, methods and media, positive-moment omissions and their causes, perceived recording effort and selectivity, and experiences of reducing or discontinuing tool use. 
For the habitual-practice reference, participants had an approximately three-day pre-deployment window to revisit their own records using usual review practices, with at least one review. They were asked to include records from the preceding few days and from approximately one to two weeks earlier. After reviewing these records, they rated \textit{Identification}, \textit{Reconstruction Detail}, \textit{Affective Re-experiencing}, and \textit{Rediscovery Awareness} on seven-point scales with reference to those review experiences (Table~\ref{tab:questionnaire-benchmark-mapping}). 
Related item sets were averaged after confirming Cronbach's $\alpha \ge 0.70$. All response-confidence ratings were at least 8 on a 0--10 scale.

\paragraph{Post-Resurfacing Questionnaire.}

Post-resurfacing questionnaires assessed participants' experiences of individual near-term and long-term reviews.
Both review types addressed the same assessment dimensions as the pre-study questionnaire, with wording specific to the current review. \textit{Rediscovery Awareness} was analyzed only for long-term reviews. Additional single-item ratings assessed \textit{Cue Sufficiency}, \textit{Laughter Entry}, \textit{Added Information}, and \textit{Personal Value}. A multiple-selection item captured the informational and experiential contributions of laughter. Participants reported whether they expanded context and whether it was needed for reconstruction or used for further exploration after recall. Near-term questions assessed external-record presence and media, deliberate-recording likelihood under usual habits, and, when an external record existed, whether laughter brought back the original emotion more strongly than that record. Appendix~\ref{app:questionnaire-operationalization} provides the items, response formats, and scoring details.

\begin{table*}[b]
\centering
\small
\begin{tabularx}{\textwidth}{
@{}
>{\raggedright\arraybackslash}p{0.24\textwidth}
>{\raggedright\arraybackslash}X
>{\raggedright\arraybackslash}p{0.18\textwidth}
@{}
}
\toprule
\textbf{Measure} &
\textbf{Shared Assessment Focus} &
\textbf{Scoring} \\
\midrule
\midrule
\textit{Identification} &
Identifying the experience to which the record corresponds. &
Single item. \\
\midrule

\textit{Reconstruction Detail} &
Events before and after the recorded moment; details distinguishing similar experiences; associated people, settings, or interactions. &
Mean of three items. \\
\midrule

\textit{Affective Re-experiencing} &
Reinstatement of the original emotion, atmosphere or connection with others, and the feeling of re-entering the scene. &
Mean of three items. \\
\midrule

\textit{Rediscovery Awareness} &
More concrete awareness of recent life, relationships, or emotional states, and a richer overall understanding of life. &
Mean of two items. \\

\bottomrule
\end{tabularx}
\caption{Shared assessment dimensions in the pre-study habitual-practice and post-resurfacing questionnaires. All items used seven-point scales. Post-resurfacing Rediscovery Awareness was analyzed only for long-term resurfacing.}
\label{tab:questionnaire-benchmark-mapping}
\end{table*}

\paragraph{Standardized Measures.}
Participants completed the \textit{World Health Organization-Five Well-Being Index} (WHO-5)~\cite{topp20155} at baseline, mid-deployment, and study completion. We interpreted these scores alongside qualitative interview findings to explore well-being trajectories. Participants also completed the standard \textit{System Usability Scale} (SUS)~\cite{brooke1996sus} at mid-deployment and study completion to assess perceived system usability.

\subsubsection{Post-Study Interviews}
At study completion, we conducted semi-structured interviews informed by each participant's pre-study practices, questionnaire responses, and selected resurfaced moments. Aligned with these four aspects of use, the interviews examined coverage beyond existing recording practices, the affective contribution of laughter, experiences of reconstruction and reflection, and perceived changes in attention to positive experiences or in well-being. The full interview guide is provided in Appendix~\ref{app:post-study-interview}.

\subsubsection{Application usage logs}
The application recorded timestamped events for capture, detail-page access, media playback, context expansion, curation, and notifications. Participant, session, moment, and visit identifiers supported analyses of recorded activity, repeated access, contextual exploration, and technical reliability. Telemetry contained metadata and identifiers but no raw audio, photos, or video. We distinguished recording occasions, unique moments, detail-page visits, questionnaire-confirmed reviews, and participant-level summaries throughout the analysis.
We also inspected user-authored notes and contextual fields in saved Moment Cards, linking them to curation and access events using participant, session, and moment identifiers.

\subsubsection{Evaluation}

To examine these aspects of use, we combined quantitative analyses of questionnaire responses and application logs with a qualitative analysis of post-study interviews. In the Results (Section~\ref{chap:Results}), we present these complementary forms of evidence together under S1--S4.

\paragraph{Habitual-Practice Contrasts.}
We summarized questionnaire ratings and coverage proportions within participants, then averaged them with equal participant weights. 
We examined whether the reviewed moments extended beyond participants’ usual recording practices using only the 54 near-term reviews.
% Coverage analyses included only the 54 near-term reviews, because questions about existing records and the likelihood of deliberate recording were administered only after near-term resurfacing.
To contextualize deployment experiences relative to habitual practice, we then compared each participant’s pre-study habitual-practice score with their mean post-resurfacing score. 
% We summarized questionnaire ratings and coverage proportions within participants, then averaged these summaries with equal participant weights (participant-equal aggregation). Coverage analyses were restricted to the 54 near-term reviews because the relevant questions about existing records and likely deliberate recording were administered only after near-term resurfacing. For within-participant descriptive contrasts with habitual practice, we paired each participant's pre-study habitual-practice score with their mean post-resurfacing score. These contrasts provide participant-specific context for interpreting deployment experiences. 
Analyses of \textit{Identification}, \textit{Reconstruction Detail}, and \textit{Affective Re-experiencing} included both temporal distances, whereas \textit{Rediscovery Awareness} used long-term reviews only.

\paragraph{App Usage and Context Use.}
% To examine how participants reviewed moments, we first grouped repeated openings into a single access episode when they involved the same participant, moment, review mode, and calendar date. The gaps between consecutive visits no more than 30 minutes.
The app usage logs were used to examine when and how participants reviewed moments. To characterize review timing, we grouped repeated openings into a single access episode when they involved the same participant, moment, review mode, and calendar date, with no gap exceeding 30 minutes. To examine review behavior, we linked each questionnaire-reported review to its corresponding moment and access episode based on participant identity, near- or long-term condition, moment descriptions and metadata, and participant confirmation. For each linked review, we coded the use of laughter audio, surrounding audio, photos, and video. Questionnaire responses were then used to classify context expansion into three pathways: no expansion, further exploration after the moment had been recalled, and expansion to support recall.

% We analyzed app access at two levels. First, to characterize when participants accessed moments, we combined repeated openings into a single access episode when they involved the same participant, moment, review mode, and calendar date, with no more than 30 minutes between consecutive visits. Second, to examine behavior during questionnaire-reported reviews, we linked each questionnaire to its corresponding moment and associated access episode. We checked these links using participant identity, the reported near- or long-term condition, moment descriptions and metadata, and participant confirmations. For each linked review, we coded whether participants used laughter audio, surrounding audio, photos, or video. From questionnaire responses, we derived three context-expansion pathways: no expansion, further exploration after the moment had already been recalled, and expansion because additional context was needed for recall.

\paragraph{Well-being and Usability.}
WHO-5 was assessed at baseline, the first designated mid-deployment assessment, and study completion. To examine associations between system use and well-being, we defined review rhythm as the proportion of observed study days with at least one moment-detail visit. The proportion of detail-page visits containing a \texttt{clip\_details} expansion served as a behavioral proxy for review depth. 
We examined how review rhythm and review depth were associated with baseline-to-completion changes in WHO-5 scores.
% We related both measures to baseline-to-completion changes in WHO-5. 
SUS scores were summarized across participants.

\paragraph{Statistical Testing.}
% Paired comparisons used two-sided Wilcoxon signed-rank tests with $p$ values obtained by enumerating all sign allocations of nonzero paired differences. Tied absolute differences received average ranks. We report mean paired differences and rank-biserial effect sizes. For participant-equal estimates and mean paired differences, 95\% percentile confidence intervals were calculated from bootstrap resamples with replacement. Exploratory associations used Spearman's rank correlation with two-sided Monte Carlo $p$ values based on 1,000,000 label permutations. Item-level analyses, temporal-distance comparisons, and participant-level associations were exploratory.
Paired comparisons used two-sided Wilcoxon signed-rank tests. We report mean differences and rank-biserial effect sizes. We calculated 95\% bootstrap percentile confidence intervals for mean differences and participant-equal estimates. Exploratory associations were assessed using Spearman correlations with two-sided Monte Carlo $p$ values based on $1,000,000$ permutations. Item-level, temporal-distance, and participant-level analyses were considered exploratory.

\paragraph{Qualitative Analysis.}
We conducted a descriptive qualitative analysis of the post-study interview records. The first two authors coded the records and discussed their coding and interpretations to develop the qualitative findings. We organized these findings around S1--S4, comparing recurring and contrasting accounts within and across participants. Interview accounts were interpreted alongside participants' pre-study practices, questionnaire responses, and usage patterns to understand their experiences with \textit{LaughAnchor} and their attributions.

\begin{figure}[t]
    \centering
    \includegraphics[width=\linewidth]{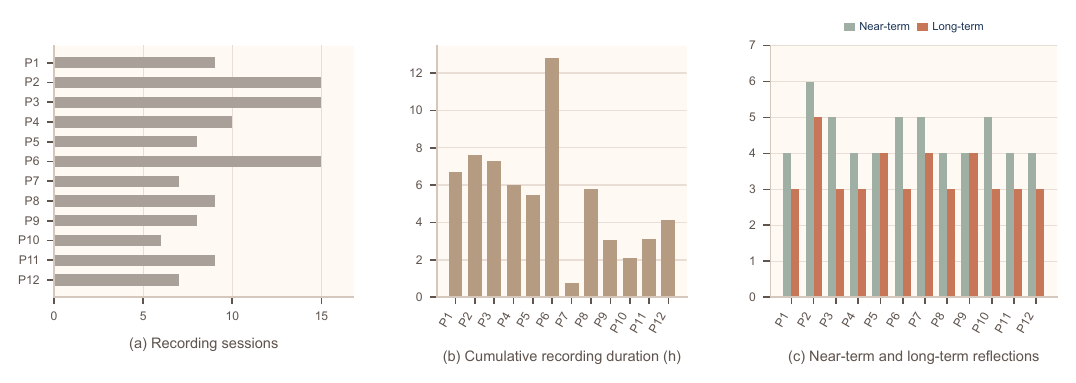}
    \caption{Participant-level recording exposure and questionnaire-confirmed reviews in the field study ($N=12$). Panels show (a) the number of recording sessions, (b) cumulative recording duration in hours, and (c) the numbers of near-term and long-term reviews linked to completed post-resurfacing questionnaires.}
    \label{fig:study_usage}
\end{figure}

\subsection{Results}
\label{chap:Results}

Across the 12 participants, the deployment collected 118 recording sessions and generated 129 candidate moments, 14 of which were deleted. Of the remaining 115 moments, 101 were opened at least once.
% Across 12 participants, the deployment collected 118 recording sessions, generated 129 candidate moments, of which 14 were deleted. 101 of 115 were opened at least once. 
We analyzed 95 questionnaire-confirmed reviews (54 near-term and 41 long-term) covering 67 distinct moments. 
There are 27 moments appeared in multiple questionnaire-confirmed reviews, including 26 for twice and one for three times. For moments reviewed at both temporal distances, the near-term and long-term reviews were separated by more than seven days.
Figure~\ref{fig:study_usage} presents the participant-level recording frequency, cumulative recording duration, and the numbers of questionnaire-confirmed reviews. All participants met the recommended minimum counts for recording and both review types, and all 95 reviews were included in the linked media-use analysis. Application logs yielded 199 identifiable user-initiated access episodes for the timing analysis.

% Across 12 participants, the deployment included 118 recording sessions. Figure~\ref{fig:study_usage} summarizes participant-level recording frequency, cumulative recording duration, and questionnaire-confirmed reviews. 
% The system generated 129 candidate moments, of which 101 were opened at least once. The 95 questionnaire-reported reviews concerned 67 distinct Moments.
% We analyzed 95 post-resurfacing questionnaires, comprising 54 near-term reviews and 41 long-term reviews. All participants met the recommended minimum counts for recording and both review types. All 95 reviews were
% included in the linked media-use analysis. Application logs yielded 199 identifiable user-initiated access episodes for the timing analysis.

\subsubsection{S1. Passive laughter indexing broadened participant-attributed coverage beyond deliberate recording.}

% Across different recording frequencies, passive laughter indexing complemented existing practices by preserving candidate traces of moments participants considered unlikely to record deliberately but later regarded as worth preserving.

\paragraph{Quantitative Results:}

%先讲用了哪几条问题分的组，分了几个组
%然后再讲每个组的情况
From the post-resurfacing questionnaire responses, we used coverage and external-record questions to divide the near-term reviews into 4 main categories. When the participant selected 'no external record', if the rated \textit{Would\_Record} (i.e., deliberate recording likelihood without the system) $\leq 3$, and \textit{Personal value} $\geq 5$, we defined as the first category with participant-attributed \textit{"valued-novel"}. Otherwise, if \textit{Would\_Record} $>3$, then defined as the second category, while those with \textit{Would\_Record} $\leq 3$ and \textit{Personal Value} $<5$ formed the third category. Reviews involving an external record formed the fourth category. For these reviews, participants additionally rated whether laughter brought back the original emotion more strongly than that record. Table~\ref{tab:near-term-classification} summarizes the resulting classification.

Among 54 near-term reviews, 24 (44.4\%), involving 23 distinct moments, met the \textit{valued-novel} criteria. Another 13 reviews (24.1\%) had no external record but received \textit{Would\_Record} ratings above 3. These categories show that \textit{LaughAnchor} complemented existing practices both by preserving valued moments participants considered unlikely to record deliberately and by capturing moments that remained unrecorded even though participants considered them more likely to document. The participant-equal valued-novel coverage was 43.6\% (95\% CI [25.0\%, 63.1\%]), and 9 of 12 participants encountered at least one such moment. In an exploratory analysis, greater reported in-the-moment recording burden was associated with a higher valued-novel proportion ($\rho=.643$, $p=.027$), consistent with capture effort contributing to missed moments.

% In the pre-study questionnaire, 8 participants reported often missing positive everyday moments. From the post-resurfacing questionnaire responses, we defined a participant-attributed \textit{"valued-novel"} moment as one with no external record, a rating $\leq 3$ for deliberate recording likelihood without the system (\textit{Would\_Record}), and a rating $\geq 5$ for \textit{Personal value}, both on seven-point scales. 
% Among 54 near-term reviews, 24 (44.4\%), involving 23 distinct moments, met these criteria (Table~\ref{tab:near-term-classification}). Another 13 (24.1\%) involved moments with no external record but recording-likelihood ratings $> 3$. Thus, the system captured both valued moments considered unlikely to be recorded and unrecorded moments rated as more likely to be documented under usual practices.
% Participant-equal valued-novel coverage was 43.6\% (95\% CI [25.0\%, 63.1\%]), and 9 of 12 participants encountered at least one such moment. In an exploratory analysis, greater reported in-the-moment recording burden was associated with higher valued-novel coverage ($\rho=.643$, $p=.027$), consistent with capture effort as one source of missed coverage. 

\begin{table}[t]
\centering
\small
\begin{tabularx}{\columnwidth}{@{}Xrrr@{}}
\toprule
\textbf{Category of Moments }& \textbf{\# of Reviews }& \textbf{\# of Moments }& \textbf{\% of Reviews} \\
\midrule
\midrule
Category 1: Valued-novel with no external record,
\textit{Would\_Record} $\leq 3$,
\textit{Personal Value} $\geq 5$
& 24 & 23 & 44.4 \\

\addlinespace
Category 2: No external record, but
\textit{Would\_Record} $> 3$
& 13 & 12 & 24.1 \\

\addlinespace
Category 3: No external record,
\textit{Would\_Record} $\leq 3$, but
\textit{Personal Value} $< 5$
& 8 & 8 & 14.8 \\

\addlinespace
Category 4: Had an external record
& 9 & 8 & 16.7 \\

\hspace*{1em}Comparative emotion-recall rating of 7
& 7 & 6 & 13.0 \\

\hspace*{1em}Comparative emotion-recall rating of 6
& 1 & 1 & 1.9 \\

\hspace*{1em}Comparative emotion-recall rating of 5
& 1 & 1 & 1.9 \\

\midrule
\textbf{Total} & \textbf{54} & \textbf{51} & \textbf{100.0} \\
\bottomrule
\end{tabularx}

\caption{Distribution of near-term review occasions across moment categories, based on responses from the coverage and external-record questions of the Post-Resurfacing Questionnaire. Near-term review occasions grouped by whether the moment had an external record, whether participants would otherwise have recorded it, and its personal value. The four main categories are mutually exclusive. Indented rows show whether laughter brought back the original emotion more strongly than the external record. The 54 reviews represent 51 distinct Moment Cards, and percentages use review occasions as the denominator.
}
\label{tab:near-term-classification}
\end{table}

\paragraph{Qualitative Results:}

% moment 的价值没有在发生当下变得足够明显。它可能显得普通，也可能发生在持续展开且使人投入的互动中，因此参与者当时没有把它识别为一个需要记录的对象。价值是在之后回顾时才显现出来。

\textbf{Capturing potential moments before retrospective value recognition.}
In post-study interviews, all participants identified at least one system-captured moment they would rarely have recorded themselves but later considered worth preserving. They attributed this additional coverage to passive indexing of moments that seemed too ordinary to document or unfolded while they were absorbed in an interaction. Participants did not need to deliberately seek interesting experiences, and the system captured \textit{"things I might initially have considered unimportant, but which felt meaningful when I reviewed them"} (P2). The accumulated traces also made the frequency of laughter visible, for instance, P5 remarked \textit{"I never imagined that I laughed this often".} \textit{LaughAnchor} preserved lightweight traces before recording intentions had formed, making these experiences available for retrospective recognition and evaluation without requiring an in-the-moment recording action.

% 对低习惯参与者，主要问题是没有时间、注意力、方法或意愿在当下发起记录，因此整个体验可能没有任何 deliberate record。
% 对高习惯参与者，问题是既有记录通常保存已经显得重要的离散节点，例如餐食、合照、地点和活动终点；持续展开的交谈、语气和笑点则位于这些节点之间，较少进入个人档案。
% 我们的system通过 xxx，对两类人

\textbf{Addressing different coverage gaps across recording practices.}
The contribution of \textit{LaughAnchor} varied with participants' existing recording practices. Pre-study accounts from lower-frequency recorders described interruption, uncertainty about how to record, and limited attention as barriers to initiating capture. For these participants, passive laughter indexing allowed an entire positive experience to enter the candidate set without requiring deliberate action during the moment. For P12, \textit{LaughAnchor} preserved a concrete trace of a cycling trip that would otherwise have left no deliberate record beyond \textit{"merely a general impression that the experience had been enjoyable".}

For higher-frequency recorders, \textit{LaughAnchor} supplemented photos, written records, and social media posts that preserved deliberately selected landmarks but often omitted intervening interactions.
P9 explained that he might photograph the arrival of food or the end of a meal but would not keep a camera or phone raised to record the intervening conversations, often \textit{"the more interesting and meaningful part"}. By indexing laughter during these undocumented intervals, \textit{LaughAnchor} extended coverage from discrete landmarks to the ongoing social interactions around them.

\begin{figure}[t]
    \centering
    \includegraphics[width=\linewidth]{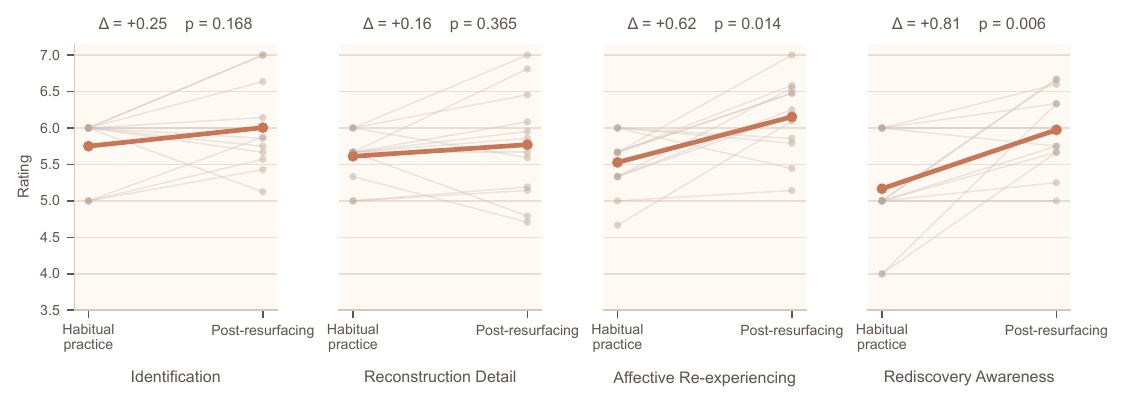}
    \caption{Within-participant descriptive contrasts between pre-study habitual-practice ratings and post-resurfacing ratings. Thin gray lines connect each participant's habitual-practice reference score and mean post-resurfacing score, while thick orange lines show participant-equal means. Panel annotations report mean paired differences ($\Delta$, post-resurfacing minus habitual-practice reference) and $p$ values.}
    \label{fig:planned-comparisons}
\end{figure}

\subsubsection{S2. Laughter provided an affective route into resurfaced experiences.}

\paragraph{Quantitative Results:}

Across 95 reviews, participants reported strong affective reconnection, with a participant-equal \textit{Affective Re-experiencing} mean of 6.15 out of 7 (95\% CI [5.86, 6.43]). For descriptive context, the mean of habitual-practice rating was 5.53, yielding a mean difference in participant rating of $+0.62$ points (95\% CI [0.26, 0.96], $p = .014$, $r_{\mathrm{rb}} = .782$), as shown in Figure~\ref{fig:planned-comparisons}.

Participants also attributed a role to laughter. Participant-equal means were 6.06 out of 7 for \textit{Laughter Entry} (95\% CI [5.71, 6.41]) and 5.72 for \textit{Added Information} (95\% CI [5.08, 6.26]). Participants most often credited laughter with conveying overall atmosphere (88/95), emotional intensity (86/95), the laughter trigger (67/95), and interaction (57/95). Meanwhile, nine near-term reviews from five participants concerned eight distinct moments that already had external records. For these moments, ratings of whether laughter brought back the original emotion more strongly than the existing record averaged 6.67 out of 7 (95\% CI [6.27, 7.00]). These descriptive results support perceived affective complementarity within this self-selected subset.

\begin{figure}[t]
    \centering
    \includegraphics[width=\linewidth]{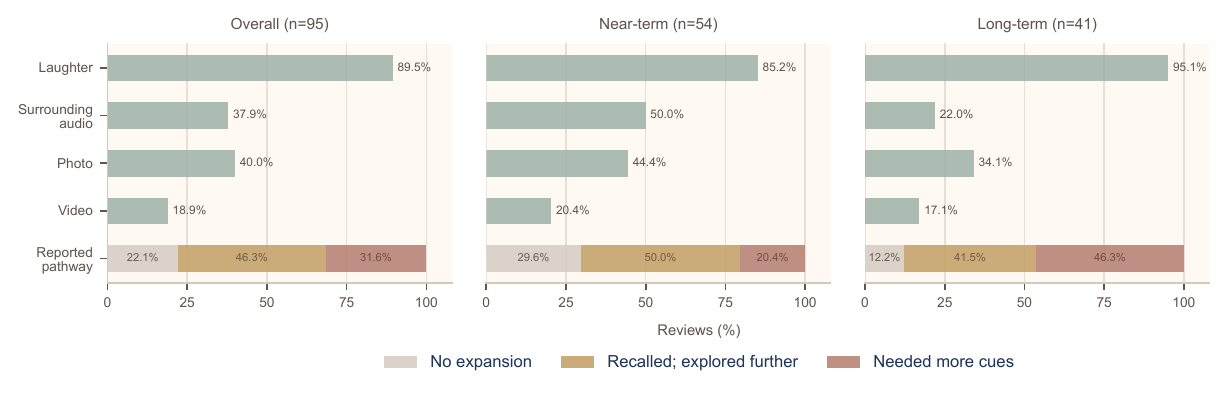}
    \caption{Media use and context-expansion pathways in questionnaire-linked reviews, overall and by temporal distance. Media-use bars show the proportion of reviews with log-recorded use of each medium. Stacked bars show three mutually exclusive questionnaire-reported pathways.}
    \label{fig:logfile_layer}
\end{figure}

Interaction logs further showed repeated engagement with laughter audio during review. Across deployment, participants initiated 1,389 laughter playbacks, of which 92.6\% reached at least half of the clip, and 77.2\% reached at least 90\%. Laughter playback occurred in 85 of the 95 questionnaire-linked reviews (Figure~\ref{fig:logfile_layer}), confirming exposure to and repeated use of laughter.

\paragraph{Qualitative Results:}

\textbf{Re-entering the emotional state of a moment.}
Participants often described laughter as rapidly bringing back the original emotional state, sometimes before they had fully articulated what had happened. P3 noted that \textit{"hearing the laughter and the bits of conversation around it makes it feel more immersive and brings me back to that moment more smoothly"}. 
P2 described laughing again while \textit{"even only listening to their own recording, without knowing what the joke even was"}.
These accounts describe affective re-experiencing beyond factual recognition of a past event.
Moreover, P5 described revisiting the same moment about a week later. \textit{"Perhaps because I had seen it before, I did not laugh out loud again"}, but she could \textit{"still relive that happiness and warmth"}. 
Saved notes also documented enjoyment during repeated review. During a near-term review, P8 wrote \textit{"This kind of content is quite funny"} and, about a week later, appended \textit{"Revisiting it is still fun and makes me happy"} to the same card. 
The later entry documented continued enjoyment when she revisited the same moment.

\textbf{Reinstating vitality and interactional qualities.}
Participants singled out laughter when explaining their affective reconnection. They specifically attributed vitality and emotional intensity to laughter, while contextual materials helped situate and interpret that response. P3 found static images less effective without sound and compared laughter to live photos that \textit{"feel full of life"}. 
Listening to the recordings also made expressive details available for comparison. P4 initially perceived little difference between recording laughter and video, but after listening to several clips noticed that \textit{"my laughter was different in different situations"}. Focusing on sound helped them attend to \textit{"how I was laughing"}. Laughter became something to examine across experiences, alongside its role in locating and re-entering an episode.

Speech accompanying laughter in the same recording played a complementary role by anchoring this affective response in the interaction. P1 described brief words interspersed with laughter as making the experience more immersive and helping her return to the moment. Although these snippets contained limited details, they situated the laughter within an unfolding exchange by indicating what was being discussed and how people responded to one another.

% \subsubsection{S3. Context-scaffolded resurfacing supported reconstruction and reflection.}
\subsubsection{S3. Contextual exploration and interpretation extended beyond episode recall.}

\paragraph{Quantitative Results:}

% We examined whether the captured laughter and contextual cues were sufficient for episode reconstruction, which was central to the design. 
Across reviews, participant-equal ratings were 6.00 out of 7 for \textit{Identification} (95\% CI [5.70, 6.34]), 5.77 for \textit{Reconstruction Detail} (95\% CI [5.38, 6.18]), and 6.02 for \textit{Cue Sufficiency} (95\% CI [5.68, 6.36]). These ratings indicate that participants generally found the records and designed mechanisms sufficient to identify episodes and recover relevant details.

% As shown in Figure~\ref{fig:logfile_layer}, in 65 of 95 reviews (68.4\%), participants did not report needing additional context for reconstruction. Of these, 21 involved no expansion, while 44, reported by 11 participants, involved further exploration after recalling the episode. Participants reported needing additional context in the remaining 30 reviews (31.6\%), including 11 of 54 near-term reviews (20.4\%) and 19 of 41 long-term reviews (46.3\%). These reports suggest that the initial presentation often supported recall, while expandable context served both reconstruction and further exploration after recall.

As shown in Figure~\ref{fig:logfile_layer}, participants did not report needing additional context for reconstruction in 65 of 95 reviews (68.4\%). In 44 of these 65 reviews (67.7\%), reported by 11 participants, they nevertheless continued exploring after recalling the episode. The remaining 21 involved no expansion. Participants reported needing additional context in the other 30 reviews (31.6\%), including 11 of 54 near-term reviews (20.4\%) and 19 of 41 long-term reviews (46.3\%). These pathways distinguish expanding context to reconstruct an episode from continuing to examine a record after the episode has been recalled.
In the descriptive contrast with habitual practice, system ratings were descriptively similar for \textit{Identification} (6.00 vs. 5.75) and \textit{Reconstruction Detail} (5.77 vs. 5.61), with no reliable paired differences (Figure~\ref{fig:planned-comparisons}).

Beyond reconstruction, the mean within-participant difference in \textit{Rediscovery Awareness} between the long-term post-resurfacing ratings and the habitual-practice reference was $+0.81$ points (95\% CI [0.38, 1.27], $p=.006$, $r_{rb}=.945$), as shown in Figure~\ref{fig:planned-comparisons}. 
This descriptive contrast highlights participants' reported awareness of recent life, relationships, and emotional states during resurfacing.

\paragraph{Qualitative Results:}

\textbf{Reconstructing episodes through laughter-indexed context.}
The participants described laughter as a retrieval anchor that directed them toward the surrounding interaction. P11 deliberately searched for \textit{"where the joke was and what we were talking about"} to return to the conversation. P12 similarly described how a cue about a person, topic, image, or sound narrowed the range of possible memories and helped gradually recover the episode. Laughter-indexed traces provided an entry point from which participants could reconstruct the laughter trigger, nearby conversation, and unfolding interaction. 
% These accounts illustrate how the lightweight representation supported episode reconstruction despite containing less participant-authored information than deliberate records.
Participants could also add their own accounts of an episode. During a long-term review, P1 added a note explaining how she had mistakenly brought a friend's return parcel indoors, leaving the courier unable to find it. The saved card thus included her account of how the incident unfolded alongside the captured cues.

\textbf{Resurfacing experiences as memories faded.}
Participants described resurfacing records both to inspect what had been captured and to recover details they no longer readily remembered. 
During near-term resurfacing, P11 checked the recorded material against a still-accessible memory. P12 explained, \textit{"I knew what had happened and what we had talked about, but I wanted to see what the app had actually recorded."}
During long-term resurfacing, P11 recovered details that were no longer readily available, and P12 described the experience as \textit{"real reviewing"} once the specific content was nearly forgotten.
Participants also described more reliance on contextual cues during long-term than near-term reviews. Recent episodes could often be recognized from the first laughter clip, as P4 said \textit{"I could roughly tell what it was"}. Later reviews involved \textit{"hearing the sound, then seeing the photos and other contexts"} (P8) to recover less accessible details. The layered presentation supported differing needs through readily available cues for recognition and expandable contextual material for more detailed reconstruction.

\textbf{Recognizing patterns and deepening personal meaning.}
Considering multiple resurfaced moments together helped participants recognize connections across experiences. P1 noticed the same friend in many happy moments and came to appreciate the relationship more. They described this appreciation as accumulating across near-term and long-term reviews, without each review necessarily producing a new realization. P7 used recurring records to identify which activities had occupied their time recently and which situations felt most relaxing. Reflection therefore extended beyond recalling individual events to recognizing recurring people, activities, and emotional states in everyday life. For P8, resurfacing also supported a developing appreciation of ordinary family conversations. These interactions came to be recognized as warm and personally meaningful, and returning to them at different times could deepen their understanding of family relationships. 
Personal significance was also articulated in user-authored notes. P5 wrote that \textit{"a streamer I had followed for years was finally engaged"}, connecting the recorded scene to a longstanding personal interest. Her addition expressed why the scene mattered to her, alongside what had happened.
Across these accounts, awareness involved recognizing patterns across moments, while meaning developed through interpreting the personal significance of those experiences.

\subsubsection{S4: Exploratory well-being changes and participants' interpretations.}\leavevmode\par
\vspace*{\baselineskip}

% \textit{LaughAnchor} had the potential to support attribution-aware self-observation by sustaining users' attention to positive experiences and helping them recognize links between everyday circumstances and fluctuations in their well-being. WHO-5 scores increased overall, but this change was not significantly associated with reflection rhythm or depth. Participants discussed both system use and everyday circumstances when explaining perceived changes.

Participants reported greater attention to positive everyday experiences and offered differing explanations for perceived emotional changes. WHO-5 scores increased overall during deployment, but these changes were not significantly associated with review rhythm or depth.

\paragraph{Quantitative Results:}
Because the WHO-5 assesses well-being over the preceding two weeks, the baseline, mid-deployment, and completion measures provided three snapshots of an exploratory trajectory. Participant-equal mean scores increased from 56.7 at baseline to 74.3 at mid-deployment and 84.3 at study completion. The baseline-to-completion increase was statistically significant ($\Delta = +27.7$, 95\% CI [12.33, 42.67], $p = .011$). 
Neither review rhythm nor review depth was significantly associated with baseline-to-completion changes in WHO-5. We treated these changes as hypothesis-generating observations, with interviews providing context for participants’ interpretations of their emotional experiences and everyday circumstances.

\paragraph{Qualitative Results:}
Most participants reported everyday circumstances broadly comparable to those of the preceding month. Only P12 reported a substantial increase in work pressure. 
In interviews, participants described how \textit{LaughAnchor} made positive experiences more salient.
P8 described the records as \textit{"evidence and traces to follow instead of relying on memory alone"}, while P3 realized that \textit{"there are many meaningful things in life, whether large or small"}. 
Participants described increased attention to small positive moments alongside stronger immediate emotional responses. P1 remarked that \textit{"there are many beautiful things around me"} and felt able to remain in \textit{"a very happy state"}, while P2 reported that \textit{"my mood has continued to improve over these weeks"}. These accounts suggest a potential role for \textit{LaughAnchor} in supporting positive attention, self-awareness, and immediate emotional reinforcement. Some participants also associated perceived improvements in mood over the study period with continued system use.

Moreover, P12 described positive emotional reinforcement during system use despite increased work pressure. They also reported that \textit{LaughAnchor} helped them recognize well-being fluctuations in relation to their everyday circumstances. This account illustrates how positive feelings during system use could coexist with broader changes in well-being linked to work pressure.

\subsubsection{Usability and Daily Fit}
\leavevmode\par
\vspace*{\baselineskip}

\begin{figure}[t]
    \centering
    \includegraphics[width=\linewidth]{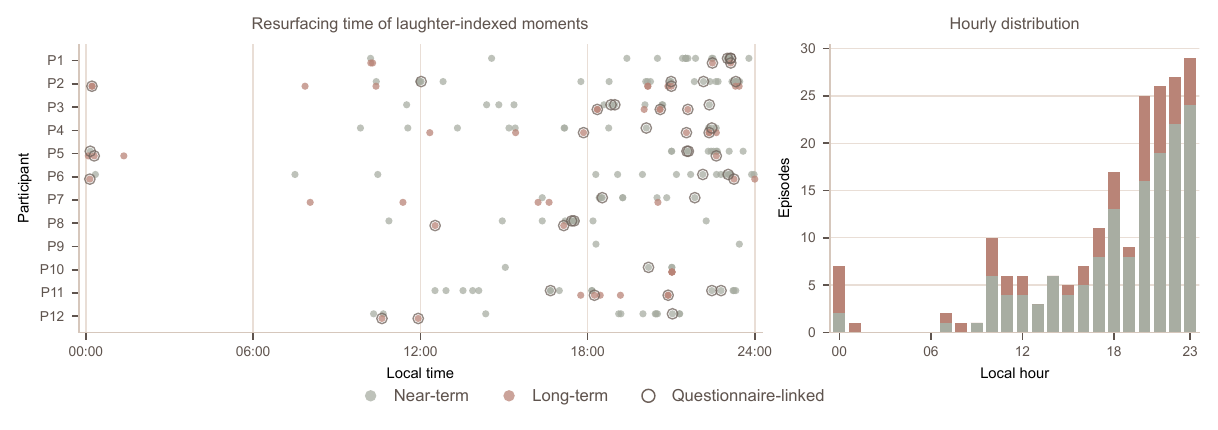}
    \caption{Local-time distribution of participant-initiated access to laughter-indexed moments. Each point in the left panel represents a logged review episode, with outlined points indicating linkage to a post-resurfacing questionnaire. Colors distinguish near-term and long-term reviews. The right panel shows hourly episode counts pooled across participants and study days.}
    \label{fig:review_timing_scatter}
\end{figure}

\textit{LaughAnchor} received a favorable SUS score at study completion ($M=83.5$, $SD=8.4$), indicating good perceived usability~\cite{lewis2018system}. Behavioral logs characterized when participants accessed retained moments. During the deployment, participants accepted 98 of the 111 resurfacing notifications (88.3\%). Across 199 identifiable user-initiated review episodes, access was concentrated in the evening, although timing varied across participants, as shown in Figure~\ref{fig:review_timing_scatter}.

Participants particularly valued the convenience of passive capture. P2 described it as \textit{"much easier than keeping a diary"}. P4 similarly appreciated that recording could remain active during daily activities without requiring them to take out a phone. These accounts suggest that passive capture accommodated ongoing activities with limited additional recording effort.
P6 and P7 also expressed interest in continuing to use the system to become more aware of positive everyday experiences. Their interest suggests that resurfacing offered a reason for future use, complementing the convenience of capture.
\section{Discussion}

\subsection{Affective Value and Contextual Exploration Beyond Precise Reconstruction}

Our findings suggest that precise reconstruction need not precede affective engagement, and recalling an episode need not end contextual exploration. Prior work has highlighted the balance between attending to laughter itself and retaining context for later recall~\cite{ryokai2018capturing}, as well as the importance of personal and interactional context~\cite{yang2022exploring}. By connecting laughter to accessible contextual materials, our deployment examines how an affective index supports the resurfacing of an experience, the reconstruction of its details, and exploratory reflection beyond recall. 

\textit{Incompleteness} could also invite exploration. P2 compared encountering such a clip to \textit{"finding an unexplained \$100 in your bank account"}, while P7 described curiosity about the surrounding context. These accounts resonate with the information-gap perspective of \citet{loewenstein1994psychology}, in which a perceived gap in understanding can motivate inquiry. Long-term resurfacing also offered unexpected rediscovery. P11 compared it to \textit{"reaching into your pocket and finding five dollars you forgot you had"}, echoing work on serendipitous encounters with personal media~\cite{helmes2011meerkat}.

Contextual exploration was not limited to occasions when participants struggled to recall an episode. In the field study, participants reported needing additional context for reconstruction in 30 of 95 reviews, while 44 involved further exploration after the episode had already been recalled. More expansion does not necessarily indicate inadequate initial cues, and successful recall does not make the remaining material redundant.

Systems should therefore keep contextual materials available for both reconstruction and continued exploration, while allowing users to stop when their purposes have been met. 
Evaluations should consider reconstruction richness, users' reasons for exploring, and the value they find in the experience. A sparse affective trace can provide an entry point into an experience whose context remains accessible.
Its affective qualities can invite engagement, while aligned contextual cues support reconstruction and further interpretation. For affective self-tracking, the design task is therefore to connect these roles without making complete recall a prerequisite for engagement or an endpoint for exploration.

% Systems should therefore keep contextual materials available for both reconstruction and continued exploration, while allowing users to stop when their purposes have been met. Evaluations should consider the reasons for expansion and the value participants derive from it alongside reconstruction richness. These findings suggest that affective resurfacing need not aim for exhaustive reconstruction of the past. Selectively preserving affective traces and leaving room for user-led contextual discovery may support a more exploratory and intriguing way of revisiting positive everyday experiences.

\subsection{From Captured Positive Moments to Everyday Awareness in Self-Tracking}

% The value of technology-assisted self-tracking extends beyond storing contextual information. 
Accumulated laughter-indexed moments can provide concrete material for noticing and interpreting changes in everyday emotional experience.
P12, for example, described recognizing emotional changes through the records but attributed these changes to everyday circumstances, not to recording itself. These records offer a selective view of life, and their significance depends on how users relate them to their lived circumstances.

Recording may also influence attention before resurfacing. Participants described seeking enjoyable experiences to record (P4) and approaching activities more positively when aware that recording was underway (P9). These accounts suggest that passive capture need not entail passive participation. Even when capture requires little moment-by-moment effort, the prospect of preserving an experience may encourage users to notice and engage with everyday sources of enjoyment. 
% These reports suggest a possible recording effect but do not establish a causal effect on well-being.

From a design perspective, recording and resurfacing may support an everyday practice of noticing as well as revisiting experiences. P5 described becoming more attentive to everyday life and spontaneously considering whether something interesting had just happened.
This resonates with lived informatics accounts of the varied purposes of personal tracking ~\cite{epstein2015lived}.
More broadly, these accounts suggest that recording systems can shape everyday attention without prescribing behavioral goals or recommendations. Anticipating future review and encountering overlooked experiences can make ordinary moments more noticeable. The design opportunity is to support this attentiveness without turning positive experience into a performance target. Longer-term work could examine whether it becomes an enduring practice, including during periods of reduced or discontinued system use.

% From a design perspective, recording and resurfacing could scaffold habits of noticing, recalling, and reflecting on positive experiences. P5 described becoming more attentive to everyday experiences and spontaneously considering whether something interesting had just happened. Integrating these activities into everyday life resonates with lived informatics accounts of personal tracking~\cite{epstein2015lived}. These changes were reported during system use, but they motivate a longer-term design aspiration to foster reflective habits whose value extends beyond continued use of the system. Future studies could examine whether such attention and reflection persist when technological support is reduced or withdrawn.

% More broadly, this suggests a role for personal informatics beyond retrospective documentation: the practice of preserving and revisiting experiences may also shape how people attend to everyday life as it unfolds. Instead of relying exclusively on quantified feedback or explicit behavioral goals, affective self-tracking may foster reflective engagement through the anticipation and reinterpretation of personally meaningful moments. This opens a broader design question of how self-tracking technologies can support everyday awareness without prescribing what users should notice or value.

% 可能需要深化，添加一点点

\subsection{Selective Records for User-Led Interpretation}

For experience-oriented personal informatics, a useful record need not provide an exhaustive account of everyday life. \textit{LaughAnchor} made candidate episodes available through sparse indexing, while users supplied perspectives that detection alone could not establish. Users supplemented the selected moments with event explanations, personal associations, and responses to later reviews.
These observations suggest that the value of selective indexing depends on both the experiences it brings forward and the opportunities it gives users to interpret them.

This design stance builds on work that supports users in interpreting affective materials~\cite{staahl2009experiencing}. 
The broader design issue is how computational selection and personal judgment are distributed across encounters. A signal can bring an experience forward without establishing its positivity or lasting value. Contextual access lets users examine what the signal refers to, while supplementation allows their own accounts to become part of the record. 
Affective sensing can thus create opportunities for reflection while leaving judgments about an experience's personal significance with the user.

This organization may inform systems using other sparse cues linked to personal media, but the cue's experiential role remains important. In \textit{LaughAnchor}, laughter was itself replayable material. Participants could reconnect with its affective qualities or attend to how they had laughed. A physiological event marker may require an additional representation or associated media to offer comparable experiential access. The transferable consideration is therefore to preserve contextual access and user authority over interpretation, without assuming that different signals will evoke the same response. 
Future research could examine how different cue-selection and contextual-presentation strategies shape which everyday experiences users notice and revisit. Evaluation could consider the range of experiences brought into reflection and the freedom users retain to interpret their personal significance.

\subsection{Limitations and Future Work}

\paragraph{Study limitations.}

Our participants were primarily university students and working adults, limiting the range of populations and everyday contexts represented. Future research should involve more diverse participants to examine how the value of laughter-indexed resurfacing and users' contextual needs vary across backgrounds and daily routines. The three-week field deployment also provided limited insight into reconstruction and reflection at longer temporal distances. Longer deployments could examine how these processes evolve as memories become more distant and whether sustained use of \textit{LaughAnchor} complements or partially replaces habitual recording and review practices.

Our comparison with habitual practice provides an ecological reference rather than a controlled comparison. Although participants revisited their own records before deployment and rated experiences along shared dimensions, the records were not matched to LaughAnchor moments in content, recording modality, or selection process. The observed differences therefore characterize participants' perceived experiences across two recording practices. Future work could use matched episodes to isolate the contributions.

\paragraph{System limitations.}

The current prototype relies on an external camera--microphone module attached to glasses and connected to a smartphone. Although this configuration supports hands-free recording, attaching the module and managing the wired connection add setup and handling demands. Future work could explore integration into commercial smart glasses to reduce these demands and examine everyday use with a more integrated form factor.

Bystander privacy remains a concern even when records are intended for private review. P1 described considering friends' privacy when deciding whether to retain moments involving them. Rich audiovisual context can aid reconstruction while revealing information about others; user curation does not replace protections during capture and processing. Future designs could explore privacy-by-default and consent-based restoration for visual material~\cite{khawaja2026see}, alongside separate safeguards for recorded speech.

\section{Conclusion}

We presented \textit{LaughAnchor}, a mobile and wearable system that uses laughter as a sparse affective index to construct contextualized personal records for reconstruction and reflection. Informed by a formative study and examined through a three-week field deployment, the system connects laughter-indexed capture with contextual grounding, user curation, and later resurfacing. Laughter provided an affective entry point, while contextual cues supported both episode reconstruction and continued exploration after recall. These findings highlight that the value of a personal record can emerge after capture and need not depend on complete recall. Resurfacing can offer enjoyment, spark curiosity, and prompt reconsideration of ordinary experiences. Designing for these possibilities involves more than making records retrievable. It means allowing people to decide how much to reconstruct, when to explore further, and what significance these moments hold in their everyday lives.

\bibliographystyle{ACM-Reference-Format}
\bibliography{ref}

\newpage
\appendix
\appendix

\section{Implementation Details}
\label{app:implementation}

\subsection{Laughter Detector Selection}
\label{sec:appendix-detector-selection}
% 比较的 detector
% 使用的公开数据集(数据量、场景和标注方式)
% 本地模型运行的 CPU、GPU
% precision、recall、F1；
% latency 或 real-time factor；
% 为什么最终选择 Speechmatics。

We benchmarked four open-source laughter detection models and selected API to inform the choice of detection backend for LaughAnchor.

\paragraph{Candidate detectors.}
Table~\ref{tab:appendix-detectors} summarizes the architectures, model size, and mechanisms of our candidate laughter detectors.
\begin{table}[h]
\centering
\small
\begin{tabular}{
|>{\raggedright\arraybackslash}p{0.10\linewidth}
|>{\raggedright\arraybackslash}p{0.15\linewidth}
|>{\centering\arraybackslash}p{0.11\linewidth}
|>{\raggedright\arraybackslash}p{0.50\linewidth}|
}
\hline
\textbf{Method} &
\textbf{Architecture} &
\textbf{Model Size} &
\textbf{Mechanism} \\
\hline
\href{https://github.com/jrgillick/laughter-detection}{Gillick}
&
ResNet
&
9.35\,MB
&
Frame-level prediction with low-pass filtering and instance detection
\\
\hline
\href{https://github.com/omine-me/LaughterSegmentation}{Omine}
&
Wav2Vec2
&
1203.36\,MB
&
Frame-level classification; trained with synthetic laughter augmentation
\\
\hline
\href{https://github.com/ideo/LaughDetection}{Ideo}
&
LSTM
&
1.53\,MB
&
Classifies fixed 3-second audio segments
\\
\hline
\href{https://github.com/hhoangphuoc/SpeechLaughRecogniser}{Hhoangphuoc}
&
Wav2Vec2 + CTC (ASR)
&
1203.67\,MB
&
Detects \texttt{<laugh>} tags / all-caps tokens in the ASR transcript
\\
\hline
\href{https://docs.speechmatics.com/}{Speechmatics}
&
Commercial API
&
---
&
Native laughter-event tag with timestamp and per-event confidence; supports real-time streaming
\\
\hline
\end{tabular}
\caption{Laughter detectors compared.}
\label{tab:appendix-detectors}
\end{table}

\paragraph{Evaluation datasets.}
We used two public datasets with complementary acoustic characteristics for benchmarking. \textit{Switchboard} primarily represents spontaneous conversational speech, whereas the \textit{AudioSet} subset provides a balanced collection of laughter and non-laughter vocal sounds drawn from online videos. Together, they allow us to evaluate laughter detection precision and recall under different acoustic conditions, reflecting both false detections and missed laughter events. Table~\ref{tab:appendix-datasets} summarizes their annotation granularity and data splits.

\begin{table}[h]
\centering
\small
\begin{tabular}{
|>{\raggedright\arraybackslash}p{0.18\linewidth}
|>{\raggedright\arraybackslash}p{0.20\linewidth}
|>{\centering\arraybackslash}p{0.07\linewidth}
|>{\centering\arraybackslash}p{0.07\linewidth}
|>{\centering\arraybackslash}p{0.07\linewidth}
|>{\raggedright\arraybackslash}p{0.20\linewidth}|
}
\hline
\textbf{Dataset} &
\textbf{Annotation Granularity} &
\textbf{Train} &
\textbf{Val} &
\textbf{Test} &
\textbf{Notes} \\
\hline
\href{https://huggingface.co/datasets/hhoangphuoc/switchboard}{Switchboard Corpus}
&
Event-level (clip label)
&
25{,}630 
& 
2{,}978 
& 
7{,}110
&
Preprocessed 16\,kHz release.
\\
\hline
\href{https://github.com/ideo/LaughDetection}{AudioSet (subset)}
&
Event-level (clip label)
&
18{,}768
&
---
&
568
&
Balanced subset extracted by the \href{https://github.com/ideo/LaughDetection}{Ideo} repository.
\\
\hline
\href{https://research.google.com/audioset/download_strong.html}{AudioSet (strong-label)}
&
Timestamp-level ([start, end] within clip)
&
---
&
---
&
39
&
Subset of the evaluation set with frame-accurate laughter boundaries; used only for timestamp-level evaluation.
\\
\hline
\end{tabular}
\caption{Public datasets used for benchmarking.}
\label{tab:appendix-datasets}
\end{table}

\paragraph{Execution environment and timing.} 
All open-source models were run locally on two NVIDIA GeForce RTX 3090 GPUs, each with 24 GB memory (compute capability 8.6). Table~\ref{tab:appendix-event} reports local processing times for these methods. Speechmatics was accessed as a hosted API. Under our benchmark conditions, its measured API turnaround times were 4.186 s on Switchboard and 5.692 s on AudioSet. These measurements include network and service processing and are reported separately from local processing times.

\paragraph{Event-level performance.}
Table~\ref{tab:appendix-event} summarizes event-level performance on the evaluation datasets. The top table reports basic detection performance using precision, recall, F1, and local processing time under the selected typical thresholds. The bottom table reports the effect of applying CTC-based ASR filtering to suppress false-positive laughter detections.

\begin{table*}[t]
\centering
\small

\begin{tabular}{|l|l|c|c|c|c|c|}
\hline
\textbf{Method} & \textbf{Dataset} & \textbf{Threshold} & \textbf{Precision} & \textbf{Recall} & \textbf{F1} & \textbf{Local processing time (ms)} \\
\hline
Omine & Switchboard & 0.50 & 0.912 & \textbf{0.760} & \textbf{0.829} & 34 \\
\hline
Omine & Switchboard & 0.65 & 0.918 & 0.725 & 0.811 & 34 \\
\hline
Omine & AudioSet & 0.50 & \textbf{1.000} & 0.645 & 0.784 & 66 \\
\hline
Omine & AudioSet & 0.65 & \textbf{1.000} & 0.613 & 0.760 & 62 \\
\hline
Speechmatics & Switchboard & 0.65 & \textbf{1.000} & 0.326 & 0.492 & --- \\
\hline
Speechmatics & AudioSet & 0.65 & \textbf{1.000} & 0.516 & 0.681 & --- \\
\hline
Gillick & Switchboard & 0.50 & 0.882 & 0.639 & 0.741 & 106 \\
\hline
Gillick & Switchboard & 0.65 & 0.903 & 0.481 & 0.627 & 103 \\
\hline
Gillick & AudioSet & 0.50 & 0.947 & 0.581 & 0.720 & 346 \\
\hline
Gillick & AudioSet & 0.65 & 0.941 & 0.516 & 0.667 & 302 \\
\hline
Ideo & Switchboard & 0.50 & 0.917 & 0.094 & 0.171 & 226 \\
\hline
Ideo & Switchboard & 0.65 & 0.923 & 0.052 & 0.098 & 95 \\
\hline
Ideo & AudioSet & 0.50 & \textbf{1.000} & 0.516 & 0.681 & 683 \\
\hline
Ideo & AudioSet & 0.65 & \textbf{1.000} & 0.452 & 0.622 & 729 \\
\hline
Hhoangphuoc & Switchboard & --- & 0.985 & 0.579 & 0.730 & \textbf{30} \\
\hline
Hhoangphuoc & AudioSet & --- & \textbf{1.000} & 0.355 & 0.524 & 36 \\
\hline
\end{tabular}

\vspace{1.5em}

\begin{tabular}{|l|c|c|c|c|c|}
\hline
\textbf{Method} & \textbf{Threshold} & \textbf{CTC} & \textbf{Precision} & \textbf{Recall} & \textbf{F1} \\
\hline
Omine & 0.50 & No  & 0.912 & \textbf{0.760} & \textbf{0.829} \\
\hline
Omine & 0.50 & Yes & 0.983 & 0.511 & 0.672 \\
\hline
Omine & 0.65 & No  & 0.918 & 0.725 & 0.811 \\
\hline
Omine & 0.65 & Yes & 0.983 & 0.494 & 0.657 \\
\hline
Gillick & 0.50 & No  & 0.882 & 0.639 & 0.741 \\
\hline
Gillick & 0.50 & Yes & 0.979 & 0.403 & 0.571 \\
\hline
Gillick & 0.65 & No  & 0.903 & 0.481 & 0.627 \\
\hline
Gillick & 0.65 & Yes & 0.988 & 0.339 & 0.505 \\
\hline
Ideo & 0.50 & No  & 0.917 & 0.094 & 0.171 \\
\hline
Ideo & 0.50 & Yes & \textbf{1.000} & 0.056 & 0.106 \\
\hline
Ideo & 0.65 & No  & 0.923 & 0.052 & 0.098 \\
\hline
Ideo & 0.65 & Yes & \textbf{1.000} & 0.039 & 0.074 \\
\hline
\end{tabular}

\caption{Top: basic event-level performance without ASR filtering. Bottom: effect of ASR/CTC-based speech filtering (Switchboard). Bold indicates the best precision, recall, and F1 values; ties are bolded jointly. In the timing column, — denotes the hosted API, which is excluded from the local-processing-time comparison.}
\label{tab:appendix-event}
\end{table*}

\paragraph{Timestamp-level performance.}
Table~\ref{tab:appendix-timestamp} reports timestamp-level detection performance on the 39-clip AudioSet strong-label subset, containing 48 ground-truth laughter events. Within each clip, predicted and ground-truth laughter intervals were matched one-to-one based on temporal overlap. For each matched pair, we computed the absolute differences between the predicted and ground-truth onset and offset times, and averaged these errors across matched pairs. We additionally report the combined boundary error, defined as the sum of the mean onset and offset errors.

\begin{table}[h]
\centering
\small
\begin{tabular}{|l|c|c|c|c|c|}
\hline
\textbf{Method} & \textbf{Threshold} & \textbf{Matched / 48} & \textbf{Onset (ms)} & \textbf{Offset (ms)} & \textbf{Combined (ms)} \\
\hline
Gillick & 0.70 & 11 & 1439 & 838 & 2277 \\
\hline
Omine & 0.70 & 16 & 936 & 1260 & 2196 \\
\hline
Ideo & 0.70 & \textbf{18} & 1539 & 1257 & 2796 \\
\hline
Hhoangphuoc & --- & 14 & 2472 & 1929 & 4401 \\
\hline
Speechmatics & 0.70 & 16 & \textbf{790} & 1035 & \textbf{1825} \\
\hline
Gillick + CTC & 0.70 & 8 & 1926 & \textbf{732} & 2658 \\
\hline
Omine + CTC & 0.70 & 16 & 935 & 1260 & 2195 \\
\hline
Ideo + CTC & 0.70 & 8 & 1412 & 1854 & 3266 \\
\hline
\end{tabular}
\caption{Timestamp-level detection performance on the AudioSet strong-label subset (39 clips, 48 ground-truth laughter events). Onset and offset errors are mean absolute errors over one-to-one matched events, and combined error is their sum. Bold indicates the best matching, onset time error, offset time error and combined time error.}
\label{tab:appendix-timestamp}
\end{table}

\paragraph{Selection rationale.}

% 选择api原因
% 需要精确的timestamp与其余context info做alignment
% 部署限制
% 误检低，漏检略差但是可以接受
% laughter as index 本身就是一种opportunism
At event-level granularity (assigning a laughter label to a $\sim$10s clip), all four open-source methods achieve reasonable precision. 
At timestamp-level granularity, Speechmatics achieves the lowest mean combined boundary error (1825 ms) among matched events, while achieving a relatively high matching count (16/48). Because LaughAnchor requires reliable timestamp-level laughter boundaries to align each detected event with its surrounding context when assembling a moment card, we considered both temporal accuracy and the limited model capacity available for on-device deployment. Although Speechmatics has relatively low recall (0.326 and 0.516 on Switchboard and AudioSet, respectively), it consistently achieves high precision (1.000 on both datasets), reducing false-positive detections. This trade-off also aligns with our semi-structured pre-study interviews, where participants were more tolerant of missed laughter than false detections: false positives could create confusion and undermine their trust in the system's accuracy, whereas missed events were generally acceptable. This tolerance is also consistent with our opportunistic use of laughter as an index rather than an exhaustive record of positive moments; indeed, some participants interpreted missed events positively, as an indication that their positive experiences were too abundant to be fully captured by automated detection. Given these considerations, we selected the Speechmatics API as our detection backend. The API output format is illustrated below:

\begin{verbatim}
{
  "session_id": "3505bc2eb878497d9df7c2be7d08dcd0",
  "started_at": "2026-04-07T14:41:24.606Z",
  "source": "browser_microphone",
  "language": "en",
  "event_count": 1,
  "events": [
    {
      "message": "laughter detected",
      "start_time": 4.16,
      "end_time": 6.72,
      "confidence": 0.7662872076034546,
      "channel": null,
      "event_type": "laughter"
    }
  ]
}
\end{verbatim}

\subsection{Event Construction and Post-Processing}

% AudioEventStarted 与 AudioEventEnded 如何配对；
% confidence threshold；
% minimum duration；
% bout grouping；
% event boundary；
% formative 与 deployment 是否使用不同参数；
% 参数如何选的

\begin{table*}[h]
\centering
\small
\begin{tabular}{|l|p{0.13\linewidth}|p{0.38\linewidth}|p{0.19\linewidth}|}
\hline
\textbf{Layer} & \textbf{Context Type} & \textbf{Collection Mechanism} & \textbf{Example} \\
\hline

\begin{tabular}[c]{@{}l@{}}L0\\\layerdesc{Laughter Only}\end{tabular}
&
Audio (laughter)
&
\textbf{Automatic} --- Speechmatics returned laughter events with start time, end time, and confidence. Events with confidence $\geq$ \texttt{LAUGHTER\_CONFIDENCE\_THRESHOLD} were retained, and the resulting laughter interval was extended by \texttt{LAUGHTER\_AUDIO\_PRE\_OFFSET} before onset and \texttt{LAUGHTER\_AUDIO\_POST\_OFFSET} after offset, with boundaries clipped to the active recording window.
&
\path{clip_000013_laughter.wav}
\\
\hline

\multirow{3}{*}{\begin{tabular}[c]{@{}l@{}}L1\\\layerdesc{Metadata Context}\end{tabular}}
&
Time
&
\textbf{Automatic} --- system clock timestamps were recorded at the onset and offset of the resulting laughter event.
&
2026-05-15 16:06:05--16:19:44
\\ \cline{2-4}

&
Location
&
\textbf{Automatic} --- phone GPS coordinates were reverse-geocoded through the AMap API, queried once per event and reused within \texttt{LOCATION\_REUSE\_WINDOW}; location was reported at \texttt{LOCATION\_GRANULARITY}.
&
Beijing... (GPS 40.157505, 116.282573; $\pm$30m)
\\ \cline{2-4}

&
Weather
&
\textbf{Automatic} --- weather conditions were queried from the AMap weather API at event end and associated with the corresponding candidate episode.
&
Fog, 21\textdegree C
\\
\hline

\multirow{2}{*}{\begin{tabular}[c]{@{}l@{}}L2\\\layerdesc{Social Context}\end{tabular}}
&
Social setting
&
\textbf{Manual} --- participants reported who they were with after each session.
&
Boris / Alone / My little nephew
\\ \cline{2-4}

&
Activity Type
&
\textbf{Manual} --- participants reported the activity or event type surrounding the laughter episode after each session.
&
Conversation with Alice / Watching show in the theatre
\\
\hline

\multirow{3}{*}{\begin{tabular}[c]{@{}l@{}}L3\\\layerdesc{Audio/Visual Context}\end{tabular}}
&
Photograph
&
\textbf{Automatic} --- \texttt{PHOTO\_COUNT} photographs were captured at \texttt{PHOTO\_CAPTURE\_TIMES} relative to the laughter onset, with a maximum of one photo-capture sequence per \texttt{MEDIA\_COOLDOWN\_WINDOW}.
&
\path{event_photo_<unix_ts>.jpg}
\\ \cline{2-4}

&
Video
&
\textbf{Automatic} --- a \texttt{VIDEO\_DURATION} clip was captured following laughter onset, with a maximum of one capture per \texttt{MEDIA\_COOLDOWN\_WINDOW}.
&
\path{event_video_<unix_ts>.mp4}
\\ \cline{2-4}

&
Audio (speech context)
&
{\raggedright \textbf{Automatic} --- a \texttt{SPEECH\_CONTEXT\_CLIP\_DURATION} rolling audio clip was retained when speech activity was detected within \texttt{SPEECH\_CONTEXT\_ADJACENT\_CLIPS} adjacent clips of a laughter event.\par}
&
\path{clip_000015_possible_related_speech_context.wav}
\\
\hline

\multirow{3}{*}{\begin{tabular}[c]{@{}l@{}}L4\\\layerdesc{Semantic Context}\end{tabular}}
&
User-authored Summary
&
\textbf{Manual} --- participants reported their mood and provided a free-text summary of the laughter episode after each session.
&
\textit{I just felt relaxed in the atmosphere, joking with friends about a meme...}
\\ \cline{2-4}

&
LLM-generated Caption
&
\textbf{Generated} --- the collected context and participant-provided information were supplied together to GPT-5.5, which was prompted to describe the laughter episode in plain, everyday language using approximately 50 Chinese characters.
&
\textit{2026-05-15 16:06 in Beijing, I was chatting with friends about a meme...}
\\ \cline{2-4}

&
Transcript
&
\textbf{Generated} --- the L3 speech-context audio was transcribed and diarized using \texttt{gpt-4o-transcribe-diarize}.
&
\textit{A: ``Did you see that meme ?'' B: ``Yeah, it was so ridiculous...''}
\\
\hline

\end{tabular}
\caption{Context information used to construct the formative study's Context Scaffold.}
\label{tab:context-scaffold}
\end{table*}

The formative study used the apparatus and laughter-capture pipeline described in Section~\ref{sec:capture-pipeline}. Table~\ref{tab:context-scaffold} reports the modality-specific rules used to generate the materials.

\subsection{Context Capture Configuration for the Formative Study and Field Study}
\label{app:formative-context-configuration}

\texttt{EVENT\_CLUSTER\_INTERVAL} was set to 1200 s in both the formative study and the field study. This parameter specifies the maximum temporal gap between consecutive laughter bouts grouped into the same candidate moment. It serves as a temporal grouping heuristic for constructing candidate records, not as a detector of semantic episode boundaries. \texttt{LAUGHTER\_CONFIDENCE\_THRESHOLD} was increased from 0.60 in the formative study to 0.70 in the field study.
Table~\ref{tab:parameter-settings} summarizes the configuration parameters used in both studies.

\begin{table}[h]
\centering
\small
\begin{tabular}{|l|c|c|}
\hline
\textbf{Parameter} & \textbf{Formative Study} & \textbf{User Study} \\
\hline
\texttt{EVENT\_CLUSTER\_INTERVAL}
& 1200\,s
& 1200\,s
\\
\hline
\texttt{LAUGHTER\_CONFIDENCE\_THRESHOLD}
& 0.60
& \textbf{0.70}
\\
\hline
\texttt{LAUGHTER\_AUDIO\_PRE\_OFFSET}
& 2.5\,s
& 2.5\,s
\\
\hline
\texttt{LAUGHTER\_AUDIO\_POST\_OFFSET}
& 2.5\,s
& 2.5\,s
\\
\hline
\texttt{LOCATION\_REUSE\_WINDOW}
& 60\,s
& 60\,s
\\
\hline
\texttt{LOCATION\_GRANULARITY}
& 100\,m
& \textbf{30\,m}
\\
\hline
\texttt{PHOTO\_COUNT}
& 2
& 2
\\
\hline
\texttt{PHOTO\_CAPTURE\_TIMES}
& +1.5\,s / +3.5\,s
& +1.5\,s / +3.5\,s
\\
\hline
\texttt{MEDIA\_COOLDOWN\_WINDOW}
& 60\,s
& \textbf{90\,s}
\\
\hline
\texttt{VIDEO\_DURATION}
& 5\,s
& 5\,s
\\
\hline
\texttt{SPEECH\_CONTEXT\_CLIP\_DURATION}
& 30\,s
& 30\,s
\\
\hline
\texttt{SPEECH\_CONTEXT\_ADJACENT\_CLIPS}
& $\pm 2$
& $\pm 2$
\\
\hline
\end{tabular}
\caption{Configuration parameters used in the formative study and user study.}
\label{tab:parameter-settings}
\end{table}

\subsection{Resurfacing Scheduling Policy}
\label{resurfacing-scheduling}

% 两大自动化推送通道
% 1. 每日回顾提醒（Daily Resurfacing）
% 触发时机：每天本地时间 19:30 通过精确闹钟触发。

% 闹钟触发后，后台线程执行两个独立的推送类别：
% Near-term：选取 昨天发生的、用户确认“Save, allow resurfacing suggestions.” 的一个欢笑时刻
% Long-term：选取 7天前发生的、用户确认“Save, allow resurfacing suggestions.” 的一个欢笑时刻
% 每个类别当天只发一次，如果当天没有符合条件的时刻，不发通知（不会发空通知）。

% 2. 地理位置提醒（Location Resurfacing）
% Group eligible historical GPS points into stable place clusters.
% % Register a proximity boundary of 50 m around each cluster-center.
% % Trigger only when the user enters the boundary after eligible moment.

% 优先级选择算法
% 从符合条件的时刻中选出"最佳"的一个，采用两级排序：

% 第一级：有用户补充内容的优先（如用户写了笔记、录了语音、拍了照片、填了社交信息）
% 第二级：媒体数量多的优先（自动抓拍 + 用户补充的音频/照片/视频总数）

% 去重与防骚扰机制
% 机制	说明
% 每日去重	每个类别（short/long）每天只发一次
% 地点每日去重	同一地点簇每天只发一次
% 冷却期	任意两个位置通知之间至少间隔 2 小时
% 事件年龄	位置提醒只对 6 小时以上的老事件生效
% 用户开关	每日提醒和位置提醒各有独立开关，默认开启

\noindent\textit{\textbf{Automated Resurfacing Channels.}} We implemented two automated channels for resurfacing eligible laughter moments:

\begin{itemize}[leftmargin=1.5em,label=--,nosep]
\item \textit{Daily Resurfacing.} A precise alarm triggers the resurfacing process at 19:30 local time each day. The background thread independently handles: (1) \textit{Near-term:} one eligible moment from the previous day; (2) \textit{Long-term:} one eligible moment from seven days prior. Both require the user to have selected ``Save, allow resurfacing suggestions.''

\item \textit{Location Resurfacing.} Eligible historical GPS points are grouped into stable place clusters. A 50-m proximity boundary is registered around each cluster center, and resurfacing is triggered when the user subsequently enters the boundary associated with an eligible moment.
\end{itemize}

\noindent\textit{\textbf{Priority Selection.}} When multiple eligible moments are available, the system selects one candidate using two-level ranking: (1) \textit{User contribution:} prioritize moments with user-added content, including notes, voice recordings, photos, or social information; (2) \textit{Media count:} among candidates at the same priority level, prioritize those with more media, including both automatically captured and user-contributed audio, photos, and videos.

\noindent\textit{\textbf{Deduplication and Notification Control.}} We applied the following constraints to limit redundant or excessive notifications:
\begin{itemize}[leftmargin=1.5em,label=--,nosep]
\item \textit{Daily deduplication:} each category (near-term/long-term) sends at most one notification per day.
\item \textit{Location deduplication:} each place cluster triggers at most one location notification per day.
\item \textit{Cooldown:} at least 2 hours between any two location notifications.
\item \textit{Event age:} location resurfacing only considers events older than 6 hours.
\item \textit{User controls:} daily and location resurfacing have independent on/off switches, both enabled by default.
\end{itemize}

\section{Scoring Procedures and Study Instruments}
\label{app:study-instruments}

\subsection{Adapted Autobiographical Interview Rubric for the Formative Study}
\label{app:scoring-procedures}

\subsubsection{Scoring Framework and Category Anchors}

To characterize the richness of participants' verbal reconstructions during the formative think-aloud study, we employed a study-specific scoring rubric informed by the Autobiographical Interview (AI) developed by \citet{levine2002aging}. The original AI distinguishes quantitative counts of internal details across five categories (event, place, time, perceptual, and thought/emotion) from qualitative ratings of episodic recollection. Its qualitative rating composite includes four 0--3 category ratings and a separate 0--6 episodic richness rating. Prior lifelogging research has also adapted autobiographical recall measures to specific evaluation objectives \cite{sas2013affectcam, niforatos2015pulsecam}.

Our adaptation retained the five internal-detail categories as dimensions of reconstruction richness but operationalized each as an ordinal rating from 0 to 3. In particular, we used 'Event' rating to characterize the specificity and coherence of participants' accounts of the laughter-related episode.

Following the general rating principles of \citet{levine2002aging}, we adopted a 0-3 scale to represent increasing levels of episodic detail richness. Table~\ref{tab:adapted-ai-rubric} provides the operational anchors for each dimension.

\begin{table}[h]
\centering
\footnotesize
\setlength{\tabcolsep}{3pt}
\renewcommand{\arraystretch}{1.25}
\begin{tabular}{|p{0.10\linewidth}|p{0.205\linewidth}|p{0.205\linewidth}|p{0.205\linewidth}|p{0.205\linewidth}|}
\hline
\textbf{Dimension} &
\textbf{0: Absent} &
\textbf{1: General} &
\textbf{2: Detailed} &
\textbf{3: Rich and Specific} \\
\hline

\textbf{Event} &
No identifiable event-specific happenings; only generic statements. &
General activity or gist of the episode, without distinctive happenings. &
Specific actions, interactions, or event triggers, but limited description of how the event unfolded. &
Rich and coherent account of the event's progression, including distinctive actions and interactions surrounding the circumstances. \\
\hline

\textbf{Place} &
No event-specific spatial information. &
Broad setting or general location. &
Identifiable venue or spatial setting with some contextual detail. &
Highly specific spatial description, including within-scene locations, arrangements, or positions. \\
\hline

\textbf{Time} &
No event-specific temporal information. &
Broad or approximate temporal reference. &
Identifiable temporal context, such as a particular time period. &
Highly specific temporal context, including precise or richly contextualized timing and temporal relationships. \\
\hline

\textbf{Perceptual} &
No event-specific sensory or perceptual information. &
General sensory impression without distinguishing detail. &
Distinct visual, auditory, tactile, or other perceptual characteristics of the episode. &
Rich, vivid perceptual description conveying distinctive sensory characteristics of the experienced scene. \\
\hline

\textbf{Thought/Emotion} &
No thoughts or emotions associated with the original episode. &
General emotional state or thought with little elaboration. &
Specific emotion, thought, or appraisal linked to an identifiable aspect of the episode. &
Richly differentiated thoughts or emotions, including their causes, development, or relationship to particular interactions. \\
\hline
\end{tabular}

\caption{Operational scoring anchors for the adapted reconstruction-richness rubric. Each dimension is rated from 0 to 3 based on the specificity and richness of the participant's verbal reconstruction.}
\label{tab:adapted-ai-rubric}
\end{table}

\subsubsection{Cumulative Scoring and Descriptive Analysis}

During the think-aloud review, each participant examined one near-term and one long-term moment, with contextual information disclosed cumulatively from L0 (laughter only) through L4 (semantic context). At each layer, ratings were assigned to the cumulative verbal account available up to and including that layer. Repeated descriptions of previously mentioned details were not treated as additional information in themselves.

For each moment and disclosure layer, the five dimension ratings were summed with equal weight to obtain a reconstruction-richness score ranging from 0 to 15:

\begin{equation}
S_{\ell} = \sum_{d \in \mathcal{D}} r_{d,\ell},
\qquad
\mathcal{D} =
\{\mathrm{Event}, \mathrm{Place}, \mathrm{Time},
\mathrm{Perceptual}, \mathrm{Thought/Emotion}\},
\end{equation}

\noindent
where $r_{d,\ell} \in \{0,1,2,3\}$ denotes the rating of dimension $d$ at disclosure layer $\ell$, and $S_{\ell} \in [0,15]$ is the corresponding cumulative reconstruction-richness score.

We examined these scores from two perspectives: 1) \textit{Cumulative richness}, where the total score at each layer characterized the richness of the participant's reconstruction up to that stage; and 2) \textit{Stage-wise changes}, where differences between successive cumulative ratings characterized changes in reconstruction richness as additional contextual information became available. For each dimension, the stage-wise change was computed as

\begin{equation}
\Delta r_{d,\ell}
=
r_{d,\ell} - r_{d,\ell-1},
\qquad \ell = 1,2,3,4.
\end{equation}

\noindent

\subsubsection{Rating and Consensus Review Procedure}

The co-first authors conducted the formative think-aloud sessions with all 12 participants, and all verbal accounts were audio-recorded and transcribed. The transcripts served as the primary material for richness scoring.
Before scoring, members of the research team were trained on the adapted five-dimensional rubric using several examples. The transcripts were then randomly assigned to the team members for initial scoring. 
The second author reviewed the initial scores against the transcripts.
Any issues identified during this review were discussed with the first author to determine the final ratings. These ratings were used in the descriptive analysis reported in Figure~\ref{fig:richness_analysis}.

\subsection{Questionnaire for Field Study}
\label{app:questionnaire-operationalization}

This section summarizes the questionnaire items underlying the reported results. Internal consistency was assessed for sets of related items using Cronbach's alpha. Item sets with $\alpha \ge 0.70$ were then combined into multi-item scores. For presentation, the constituent items are grouped under the names assigned to the resulting measures in the main text. Other ratings and categorical responses were analyzed separately. The descriptions summarize item content and do not reproduce the original questionnaire order or numbering.

\subsubsection{Pre-Study Habitual-Practice Questionnaire}

Participants were instructed to recall their practices over the preceding 30 days when answering the following background questions and individual ratings.

\begin{enumerate}
    \item How often did you deliberately record everyday life during the preceding 30 days? (Categorical frequency.)

    \item Which methods or media did you usually use to record these moments during the preceding 30 days? (Multiple selections and supplementary text.)

    \item During the preceding 30 days, how often did you leave happy, relaxing, interesting, or otherwise worth-preserving moments unrecorded? (Categorical omission frequency.)

    \item If you felt that moments had been left unrecorded, what were the main reasons? (Multiple selections and supplementary text.)

    \item How often did you deliberately revisit previous records during the preceding 30 days? (Categorical frequency.)

    \item Have you ever reduced or stopped using a recording tool or method? If so, please identify it and explain why. (Yes/no and open-ended explanation.)

    \item I often realize afterward that a valuable moment was not recorded. (Seven-point scales)

    \item Starting a recording during an experience noticeably interrupts the experience or adds effort. (Seven-point scales)

    \item I usually record only experiences that already seem important or special as they occur. (Seven-point scales)
\end{enumerate}

During the approximately three-day pre-deployment window, participants used their usual methods to review existing records from the preceding few days and from approximately one to two weeks earlier. At least one review was required, with no upper limit. They rated the following four constructs after these reviews. All items used seven-point scales.

\paragraph{\textit{Identification} (single item).}
Using my usual recording methods for these records, I can identify the experience to which a record corresponds.

\paragraph{\textit{Reconstruction Detail} (mean of three items).}
\begin{enumerate}
    \item Using my usual recording methods for these records, I can explain what happened before and after the recorded moment.

    \item Using my usual recording methods for these records, I can recall specific details that distinguish the experience from similar episodes.

    \item Using my usual recording methods for these records, I can recall the people, settings, or interactions associated with the recorded moment.
\end{enumerate}

\paragraph{\textit{Affective Re-experiencing} (mean of three items).}
\begin{enumerate}
    \item Revisiting these records through my usual practices brings back the original emotion.

    \item Revisiting these records through my usual practices brings back the original atmosphere or sense of connection with others.

    \item Revisiting these records through my usual practices makes me feel that I am re-entering the original scene, not just recognizing that it occurred.
\end{enumerate}

\paragraph{\textit{Rediscovery Awareness} (mean of two items).}

\begin{enumerate}
    \item Revisiting these records through my usual practices makes me more concretely aware of my recent life, relationships, or emotional states.

    \item Revisiting these records through my usual practices enriches my overall understanding of life around them.
    
\end{enumerate}

\subsubsection{Post-Resurfacing Questionnaire}

Participants answered the following items about the laughter-indexed moment and their current resurfacing. Rating items used seven-point scales; categorical and multiple-selection questions are identified separately.

\paragraph{\textit{Identification} (single item).}
After this review, I can clearly identify the experience to which the laughter corresponds.

\paragraph{\textit{Reconstruction Detail} (mean of three items).}
\begin{enumerate}
    \item I can explain what happened before and after the laughter.

    \item I can recall specific details that distinguish this episode from similar experiences.

    \item I can recall the people, settings, or interactions associated with this episode.
\end{enumerate}

\paragraph{\textit{Affective Re-experiencing} (mean of three items).}
\begin{enumerate}
    \item This review brought back the original emotion.

    \item This review brought back the original atmosphere or sense of connection with others.

    \item This review made me feel that I was re-entering the original scene, not just recognizing that it occurred.
\end{enumerate}

\paragraph{\textit{Cue Sufficiency} (single item).}
The cues presented during this review were sufficient to help me understand the moment.

\paragraph{\textit{Laughter Entry} (single item).}
Laughter itself provided an important entry point into this memory.

\paragraph{\textit{Added Information} (single item).}
Laughter provided information that photographs or text did not convey.

\paragraph{\textit{Personal Value} (single item).}
How personally meaningful or valuable is this moment to you now?

\paragraph{Laughter contributions (multiple selections).}
What information or experiential qualities did laughter add to this review? Categories reported in the main text included overall atmosphere, emotional intensity, the laughter trigger, and interaction. Selections were counted separately and were not averaged.

\paragraph{Context-expansion pathway (categorical self-report).}
Did you access additional context during this review? If so, was it needed to complete reconstruction, or did you continue exploring after you had already recalled the episode?

\paragraph{Coverage and external-record questions (near-term analysis).}
\begin{enumerate}
    \item Had this moment already been recorded outside LaughAnchor? (Presence or absence of an external record.) (if 'Presence', go to (3)(4); otherwise, go to (2))

    \item \textit{Would\_Record} (single item): Under your usual recording habits, how likely would you have been to deliberately record this moment?

    \item If an external record existed, which methods or media were used? (Conditional branch; multiple selections.)

    \item If an external record existed, did laughter bring back the original emotion more strongly than that record? (Conditional branch; single-item comparative rating.)
\end{enumerate}

\paragraph{\textit{Rediscovery Awareness} (mean of two items, long-term analysis).}

\begin{enumerate}
    \item Revisiting this record makes me more concretely aware of my recent life, relationships, or emotional states.

    \item Revisiting this record enriches my overall understanding of life around them.
    
\end{enumerate}

\subsection{Post-Study Interview Questions}
\label{app:post-study-interview}

The post-study interviews were semi-structured and lasted approximately 20 minutes. Before each interview, the interviewer reviewed the participant's pre-study recording practices and event-level questionnaire responses. Only the questions in Part 1 differed according to participants' prior recording practices; all participants completed Parts 2--4.

\paragraph{Part 1: Capture and Coverage.}

\textit{For participants with regular prior recording practices:}

\begin{enumerate}
    \item During the study, was there a positive moment that you probably would not have deliberately recorded but later considered worth retaining? Please describe an example and explain why you would not normally have recorded it.

    \item Did laughter-indexed capture preserve anything that your usual recording practices would not have covered, such as conversational content, emotion, atmosphere, or interactional qualities?
\end{enumerate}

\textit{For participants with limited prior recording practices:}

\begin{enumerate}
    \item What mainly explains your limited prior recording: a lack of habit, limitations of previous tools, or another reason?

    \item During the study, was there a positive moment that you probably would not have deliberately recorded but later considered worth retaining? Please describe an example. How well did the captured material represent the positive or pleasant qualities of the experience?

    \item What made you willing to retain or review this moment again? Were there also captured moments that you did not want to retain or use? Why?
\end{enumerate}

\paragraph{Part 2: Affective Contribution of Laughter.}

\begin{enumerate}
    \item Can you describe a review in which laughter or surrounding voices helped you re-experience the emotion, atmosphere, or interaction of the original moment? It is also acceptable if no review had this effect.

    \item What information or feeling, if any, did the laughter itself add? How would the review have differed if the laughter were removed but the photographs, video, text, and surrounding speech remained?

    \item Did the contribution of laughter differ between near-term reviews and long-term reviews conducted approximately one week later?
\end{enumerate}

\paragraph{Part 3: Reconstruction and Reflection.}

\begin{enumerate}
    \item Did reviewing laughter-indexed moments simply help you recall or re-experience what happened, or did any review lead to a new understanding of the event, a relationship, or your recent life? Please describe an example.

    \item How did your experience differ between near-term and long-term reviews? Did the purpose or value of reviewing a moment change over time?

    \item If a review helped you reconstruct an event but produced no new feeling or understanding, would the review still have value? Why or why not?
\end{enumerate}

\paragraph{Part 4: Everyday-Life Awareness and Wellbeing.}

\begin{enumerate}
    \item Did using the system affect how you viewed your recent everyday life? For example, did it change your awareness of positive experiences, relationships, or emotional states? It is also acceptable if you noticed no change.

    \item If you noticed a change in mood, attention, or wellbeing, what do you think contributed to it: the act of recording and reviewing, the experiences that occurred during the study period, both, or something else?

\end{enumerate}

\end{document}